\documentclass[10pt,journal,compsoc]{IEEEtran}

\usepackage[numbers]{natbib}
\usepackage{url}
\usepackage{booktabs}
\usepackage{algorithm}
\usepackage{algorithmic}
\usepackage{amssymb}
\usepackage{array}
\usepackage[normalem]{ulem}
\usepackage{footmisc}
\usepackage{graphicx}
\usepackage{epstopdf}
\usepackage{multirow}
\usepackage{enumitem}
\usepackage[skip=0pt]{caption}
\usepackage{subfigure}
\usepackage{xcolor}
\usepackage{pifont}
\usepackage{acronym}
\usepackage{amsmath}

\usepackage{hyperref}
\usepackage[hyphenbreaks]{breakurl}

\usepackage{acronym}
\acrodef{NOR}{neural outfit recommendation}
\acrodef{CNN}{convolutional neural network}
\acrodef{BPR}{Bayesian personalized ranking}
\acrodef{RNN}{recurrent neural network}
\acrodef{MLP}{multi-layered perceptron}
\acrodef{GRU}{gated recurrent unit}
\acrodef{NIL}{negative log-likelihood}
\acrodef{AUC}{Area Under the ROC curve}
\acrodef{MAP}{Mean Average Precision}
\acrodef{MRR}{Mean Reciprocal Rank}

\newcommand{\cmark}{\ding{51}}
\newcommand{\xmark}{\ding{55}}

\begin{document}

\title{Explainable Outfit Recommendation with Joint Outfit Matching and Comment Generation}

\author{Yujie Lin$^*$, Pengjie Ren$^*$, Zhumin Chen, Zhaochun Ren, Jun Ma, and Maarten de Rijke
\IEEEcompsocitemizethanks{
\IEEEcompsocthanksitem Yujie Lin, School of Computer Science and Technology, Shandong University, Jinan, China, E-mail: yu.jie.lin@outlook.com
\IEEEcompsocthanksitem Pengjie Ren, Informatics Institute, University of Amsterdam, Amsterdam, The Netherlands, E-mail: p.ren@uva.nl
\IEEEcompsocthanksitem Zhumin Chen, School of Computer Science and Technology, Shandong University, Jinan, China, E-mail: chenzhumin@sdu.edu.cn
\IEEEcompsocthanksitem Zhaochun Ren, School of Computer Science and Technology, Shandong University, Jinan, China, E-mail: zhaochun.ren@sdu.edu.cn
\IEEEcompsocthanksitem Jun Ma, School of Computer Science and Technology, Shandong University, Jinan, China, E-mail: majun@sdu.edu.cn
\IEEEcompsocthanksitem Maarten de Rijke, Informatics Institute, University of Amsterdam, Amsterdam, The Netherlands, E-mail: derijke@uva.nl
\IEEEcompsocthanksitem $^*$These authors contributed equally.
}
\thanks{}}

\markboth{}%
{}

\IEEEtitleabstractindextext{
\begin{abstract}
Most previous work on outfit recommendation focuses on designing visual features to enhance recommendations.
Existing work neglects user comments of fashion items, which have been proved to be effective in generating explanations along with better recommendation results.
We propose a novel neural network framework, \acfi{NOR}, that simultaneously provides outfit recommendations and generates abstractive comments.
\ac{NOR} consists of two parts: outfit matching and comment generation.
For outfit matching, we propose a convolutional neural network with a mutual attention mechanism to extract visual features.
The visual features are then decoded into a rating score for the matching prediction.
For abstractive comment generation, we propose a gated recurrent neural network with a cross-modality attention mechanism to transform visual features into a concise sentence.
The two parts are jointly trained based on a multi-task learning framework in an end-to-end back-propagation paradigm.
Extensive experiments conducted on an existing dataset and a collected real-world dataset show \ac{NOR} achieves significant improvements over state-of-the-art baselines for outfit recommendation.
Meanwhile, our generated comments achieve impressive ROUGE and BLEU scores in comparison to human-written comments.
The generated comments can be regarded as explanations for the recommendation results.
We release the dataset and code to facilitate future research.
\end{abstract}

\begin{IEEEkeywords}
Outfit recommendation, explainable recommendation
\end{IEEEkeywords}}

\maketitle

\IEEEdisplaynontitleabstractindextext

\IEEEpeerreviewmaketitle


\IEEEraisesectionheading
{\section{Introduction}
\label{sec:introduction}}

\begin{figure}[h]
 \centering
 \includegraphics[width=1\columnwidth]{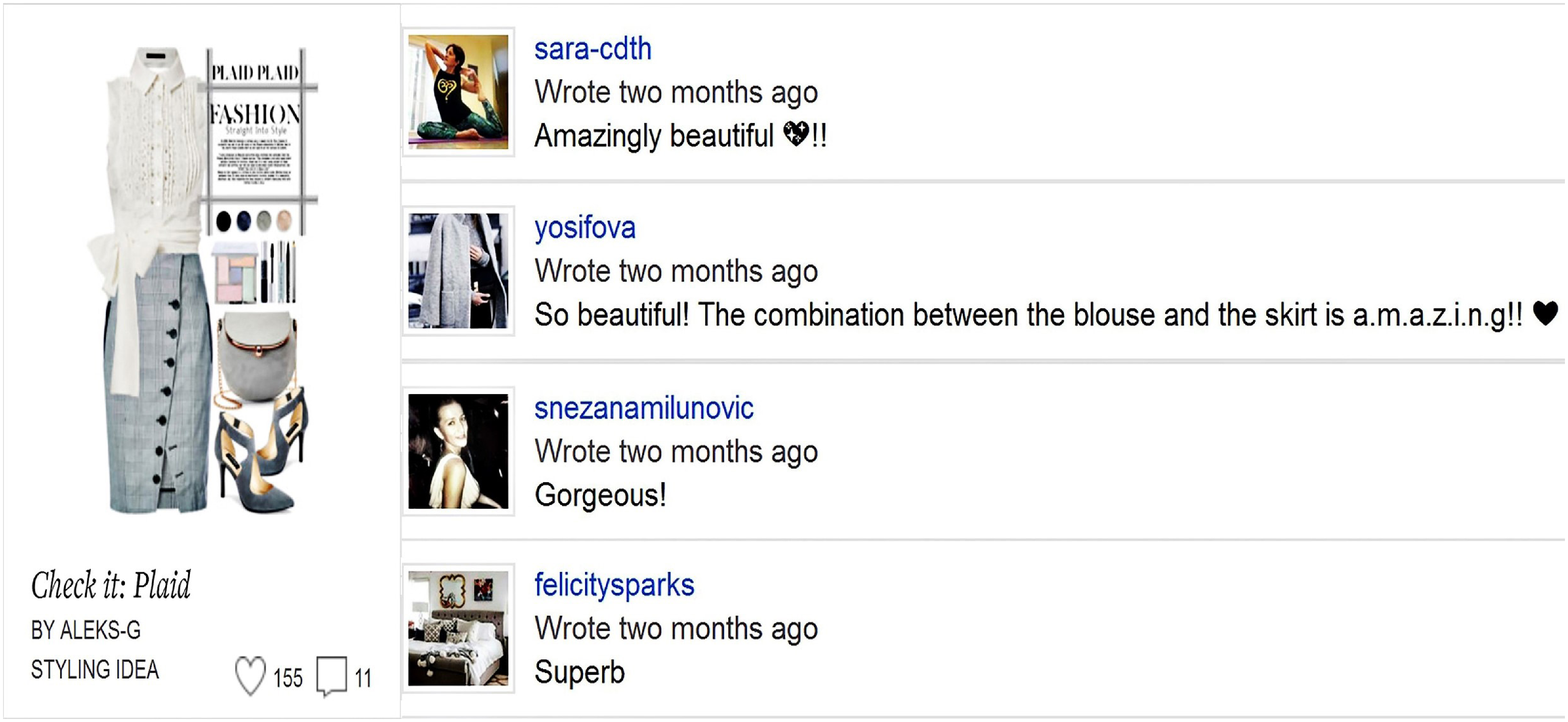}
 \includegraphics[width=1\columnwidth]{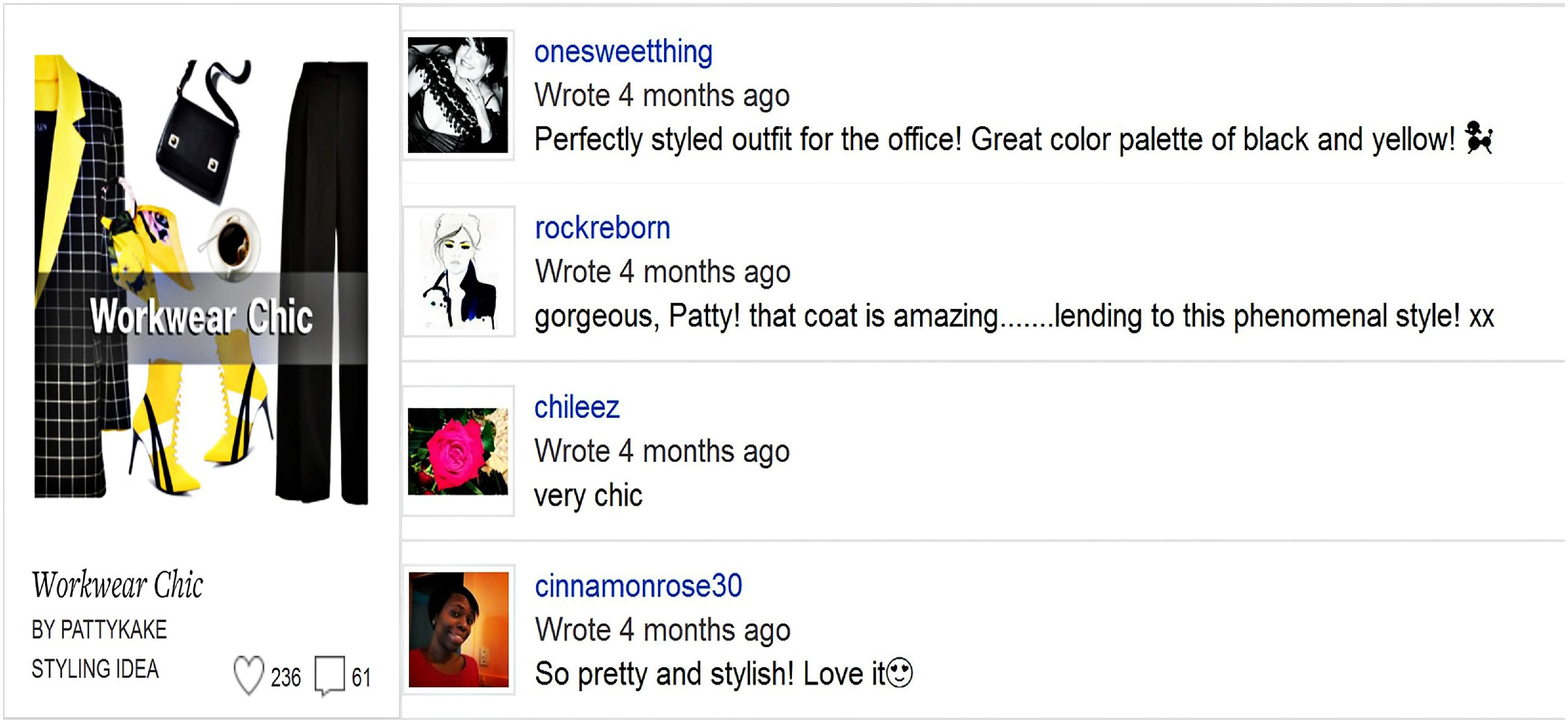}
 \caption{Outfits with user comments from Polyvore. Users share their outfit compositions with a broad public (left) and others express their comments about the outfit compositions (right).}
  \label{intro_example}
\end{figure}

\IEEEPARstart{O}{utfit} recommendation plays an increasingly important role in the online retail market.\footnote{\url{http://www.chinainternetwatch.com/19945/online-retail-2020}}
The purpose of outfit recommendation is to promote people's interest and participation in online shopping by recommending fashionable outfits that they may be interested in.
Early studies on outfit recommendation are based on small but expert-annotated datasets~\citep{Iwata2011,Liu2012}, which prohibits the development of complex models that need large sets of training material (e.g., deep learning-based models).
In recent years, with the proliferation of fashion-oriented online communities, e.g., Polyvore\footnote{\url{http://www.polyvore.com/}} and Chictopia,\footnote{\url{http://www.chictopia.com/}} people can share and comment on outfit compositions, as shown in Fig.~\ref{intro_example}.
In addition to a large number of outfit compositions, such crowdsourced data also contains valuable information (e.g., user comments) for building more accurate and intelligent recommender systems.

We address the task of explainable outfit recommendation.
Given a top (i.e., upper garment), we need to recommend a short list of bottoms (e.g., trousers or skirts) from a large collection that best match the top and meanwhile generate a sentence for each recommendation so as to explain why {the top and the bottom match}, and vice versa.
By explaining why an outfit is recommended, a recommender system becomes more transparent and trustful, which helps users make faster and better decisions~\citep{tintarev2007survey}.
The task of explainable outfit recommendation is non-trivial because of two main problems:
(1)~We need to model the compatibility of fashion factors, e.g., color, material, pattern, shape, etc.~\cite{Song2017}.
(2)~We need to model transformations between visual and textual information, which involves mappings from the visual to the textual space.

To address the problems listed above, we propose a neural multi-task learning framework, called  \acfi{NOR}.
\ac{NOR} consists of two core ingredients: outfit matching and comment generation.
For outfit matching, we employ a \ac{CNN} with a mutual attention mechanism to extract visual features of outfits.
Specifically, we first utilize \acp{CNN} to model tops and bottoms as latent vectors; then we propose a mutual attention mechanism that extracts better visual features of both tops and bottoms by employing the top vectors to match the bottom vectors, and vice versa.
The visual features are then decoded into a rating score as the matching prediction.
For abstractive comment generation, we propose a gated \ac{RNN} with a cross-modality attention mechanism to transform visual features into a concise sentence.
Specifically, for generating a word, \ac{NOR} learns a mapping between the visual and textual space, which is achieved with a cross-modality attention mechanism.
All neural parameters in the two parts of our framework as well as the word embeddings are learned by a multi-task learning approach in an end-to-end back-propagation training paradigm.

There have been several studies on outfit recommendation~\citep{Iwata2011,Liu2012,Hu2015}.
The work most similar to ours is by~\citet{Song2017}, who first employ a dual auto-encoder network to learn the latent compatibility space, where they jointly model a coherence relation between visual features (i.e., images) and contextual features (i.e., categories, tags).
Then they employ \ac{BPR}~\cite{Rendle2009} to exploit  pairwise preferences between tops and bottoms.
The differences between our work and theirs are three-fold.
First, our model can not only recommend tops and bottoms, but also generate a readable sentence as a comment.
Second, we introduce a mutual and cross-modality attention mechanism to the latent compatibility space instead of a dual auto-encoder network.
Third, we jointly train feature extraction and preference ranking in a single back-propagation scheme.

We collect a large real-world dataset from Polyvore.\footnote{\url{http://www.polyvore.com/}}
Our dataset contains multi-modal information, e.g., images, contextual metadata of items and user comments, etc.
Extensive experimental results conducted on this dataset show that \ac{NOR} achieves a better performance than state-of-the-art models on outfit recommendation, in terms of AUC, MAP, and MRR.
Moreover, comments generated from \ac{NOR} achieve impressive ROUGE and BLEU scores.

To sum up, our contributions are:

\begin{itemize}[nosep,leftmargin=10pt]
\item We explore user comments for improving outfit recommendation quality along with explanations.
\item We propose a deep learning based framework named \ac{NOR} that can simultaneously yield outfit recommendations and generate abstractive comments with good linguistic quality simulating public experience and feelings.
\item We use mutual attention to model the compatibility between fashion items and cross-modality attention to model the transformation between the visual and textual space.
\item Our proposed approach is shown to be effective in experiments on an existing dataset and a purpose-built large-scale dataset.
\end{itemize}


\section{Related Work}
\begin{figure*}
 \centering
 \subfigure{
 \includegraphics[width=1.0\textwidth]{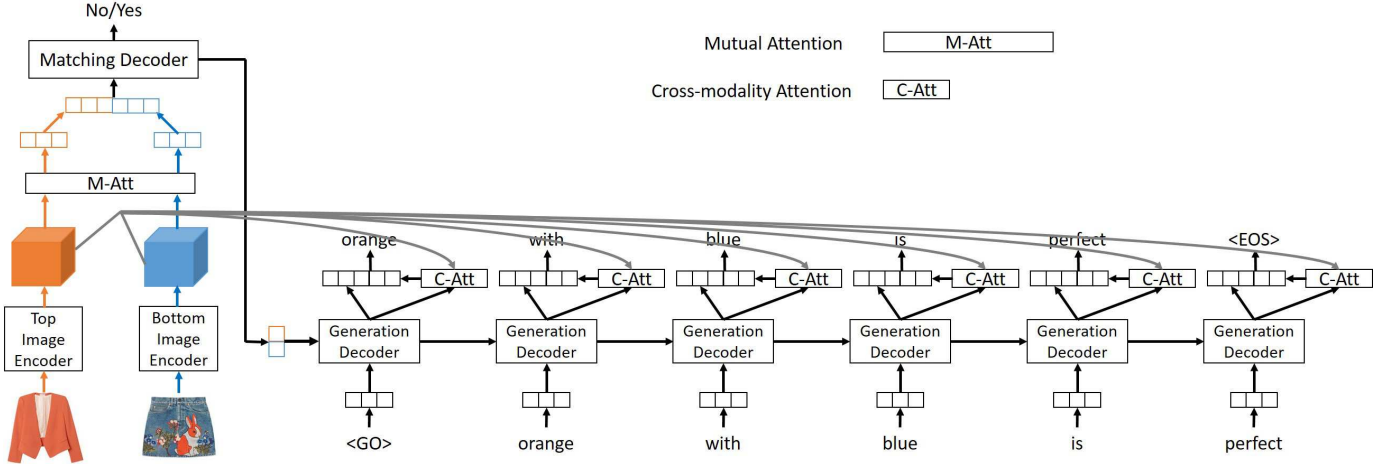}}
 \caption{Overview of the proposed \acf{NOR} architecture. \ac{NOR} contains three parts: (1)~a top and bottom image encoder (corresponding to Fig.~\ref{f_3_2a}), (2)~a matching decoder (corresponding to Fig.~\ref{f_3_2b}), and (3) a generation decoder (corresponding to Fig.~\ref{f_3_2c}).}
  \label{f_3_1}
\end{figure*}
No previous work has studied the task of explainable outfit recommendation by generating natural language comments as explanations. We briefly survey related work on outfit recommendation, on explainable recommendation and on text generation, respectively.

\subsection{Outfit recommendation}
Given a photograph of a fashion item (e.g., tops), an outfit recommender system attempts to recommend a photograph of other fashion items (e.g., bottoms).
There have been a handful of attempts to solve the task.
\citet{Iwata2011} propose a probabilistic topic model to recommend tops for bottoms by learning information about coordinates from visual features in each fashion item region.
\citet{Liu2012} study both outfit and item recommendation problems.
They propose a latent Support Vector Machine model for occasion-oriented outfit recommendation, that is, given a user-input occasion, suggesting the most suitable clothing, or
recommending items to pair with the reference clothing.
\citet{Jagadeesh2014} propose two classes of outfit recommenders, namely deterministic and stochastic, while they mainly focus on color modeling for outfit recommendation.

The studies listed above are mostly based on a small, manually annotated dataset, which prevents the development of complex models.
Several recent publications have resorted to other sources, where rich data can be harvested automatically, e.g., in the area of personalized whole outfit recommendation.
\citet{Hu2015} propose a functional tensor factorization method to model interactions between users and fashion items over a dataset collected from Polyvore.
\citet{McAuley2015} employ a general framework to model human visual preference for a pair of objects from the Amazon co-purchase dataset; they extract visual features with \acp{CNN} and introduce a similarity metric to uncover visual relationships.
Similarly, \citet{He2016} introduce a matrix factorization approach that incorporates visual signals into predictors of people's opinions.
To take contextual information (such as titles and categories) into consideration,
\citet{Li2017MiningFO} classify a given outfit as popular or non-popular through a multi-modal and multi-instance deep learning system.
To aggregate multi-modal data of fashion items and contextual information, \citet{Song2017} first employ an auto-encoder to exploit their latent compatibility space.
Then, they employ Bayesian personalized ranking to exploit pairwise preferences between tops and bottoms.
\citet{Kang2017Visually} use \acp{CNN} to learn image representations and jointly employ collaborating filtering to recommend fashion items for users.
\citet{Han2017} propose to jointly learn the visual-semantic embeddings and the compatibility relationships among fashion items in an end-to-end manner. 
They train a bidirectional LSTM model to sequentially predict the next item conditioned on previous ones to learn their compatibility relationships.
\citet{Song2018Neural} consider fashion domain knowledge for clothing matching and propose a teacher-student scheme to integrate it with neural networks. 
And they also introduce an attentive scheme to assign rule confidence in the knowledge distillation procedure.

Even though there is a growing number of studies on outfit recommendation, none of them takes user comments into account and none can give both recommendations and readable comments like we do in this paper.

\subsection{Explainable recommendation}
Explainable recommendation not only provides a ranked list of items, but also gives explanations for each recommended item.

Existing work on explainable recommendation can be classified into different categories, depending on the definition of explanation used.
Here, we only survey the most closely related studies.
\citet{Vig2009} propose an explainable recommendation method that uses community tags to generate explanations.
\citet{Zhang2014} propose an explicit factor model to predict ratings while generating feature-level explanations about why an item is or is not recommended.
\citet{He2015} propose TriRank and integrate topic models to generate latent factors for users and items for review-aware recommendation.
\citet{Ribeiro2016} propose LIME, a novel explanation technique that explains the predictions of any classifier in an interpretable and faithful manner, by learning an interpretable model locally around an individual prediction.
\citet{Ren2017} propose a richer notion of explanation called viewpoint, which is represented as a tuple of a conceptual feature, a topic and a sentiment label; though they provide explanations for recommendations, the  explanations are simple tags or extracted words or phrases.
\citet{wang2018explainable} develop a multi-task learning solution which uses a joint tensor factorization to model user preference for recommendation and opinionated content for explanation; the algorithm can generate explanations by projecting the features and opinionated phrases onto the space jointly spanned by the user and item factors.
In contrast, we generate concise sentences that express why we recommend an outfit based on all user comments. 
And we believe that simulating users to generate comments is a better way to be closer to the user's perspective which fully expresses the users' experience and feelings, making it easier for users to understand and accept.

Some recent work generates text as explanations while providing recommendations. 
\citet{NiLipVikMcA17} jointly perform personalized recommendation and review generation by combining collaborative filtering with LSTM-based generative models.
\citet{Li2017}'s work is most similar to ours.
By introducing \acp{RNN} into collaborative filtering, they jointly predict ratings and generate tips, which express the sentiment of users while reviewing an item.
Our work and previous ones differs in four ways.
First, we target a different task, i.e., we focus on outfit recommendation not score rating.
Second, the recommendation and generation in this paper are not personalized. 
We determine whether the outfit is matched based on the public perspective.
Because the factors that influence people's selections of clothes mainly include the current fashion, occupation, age and regionalism, and we believe that people with similar ages and popularity are usually similar on these factors.
Furthermore the generated comments learned from multiple comments (from different online users) reflect a general and common opinion on behalf of multiple users instead of a single specific user.
Third, unlike \citet{NiLipVikMcA17}'s work and \citet{Li2017}'s work, our task involves multiple modalities (i.e., image and text).
Fourth, instead of using a simple \ac{RNN}, we propose a more complex cross-modality attention mechanism to handle the mapping from the visual to the textual space.

\subsection{Text generation}
Text generation involves a wide variety of tasks and studies, such as text summarization~\cite{Li2017Deep,Zhou2017Selective}, machine translation~\cite{Bahdanau2014Neural,vaswani2017attention}, dialogue systems~\cite{Serban2016Building,Williams2017Hybrid}, and image captioning~\cite{Xu2015Show,Chen2017SCA}.
We list some related works on comment or review generation as follows.

\citet{Cao2012News} present a framework to automatically collect relevant microblogs from microblogging websites to generate comments for popular news on news websites.
\citet{lipton2015capturing} design a character-level \ac{RNN} to generate reviews. 
The generated reviews are based on auxiliary information, such as user/item IDs, categories and ratings.
\citet{radford2017learning} also train a character-level \ac{RNN} language model on the Amazon review dataset, which has only one single multiplicative LSTM layer with 4,096 hidden units. 
They introduce a special unit among the hidden units that can control the sentiment of the generated reviews.
\citet{Dong2017Learning} propose an attribute-to-sequence model to generate product reviews for given attribute information including user, product and ratings.
They first use an attribute encoder to learn the representation vectors of the input attributes.
Then they employ a stacked LSTM with an attention mechanism to generate reviews based on these representation vectors.
\citet{Tang2016Context} propose two novel approaches that first encode contexts, such as sentiment ratings and product ids, into a continuous semantic representation and then decode the semantic representation into reviews with \acp{RNN}. 
\citet{Hu2017Controllable} combine a variational auto-encoder and a holistic attribute discriminator to generate reviews. 
They alternately train the auto-encoder and the discriminator.
They can dynamically control the attributes of the generated reviews by learning disentangled latent representations with designated semantics.
These studies only focus on text generation, and do not jointly perform recommendation.

\section{Neural Outfit Recommendation}
\begin{figure*}
 \centering
 \subfigure[Top and bottom image encoder.]{
 \label{f_3_2a}
 \includegraphics[width=0.3\textwidth]{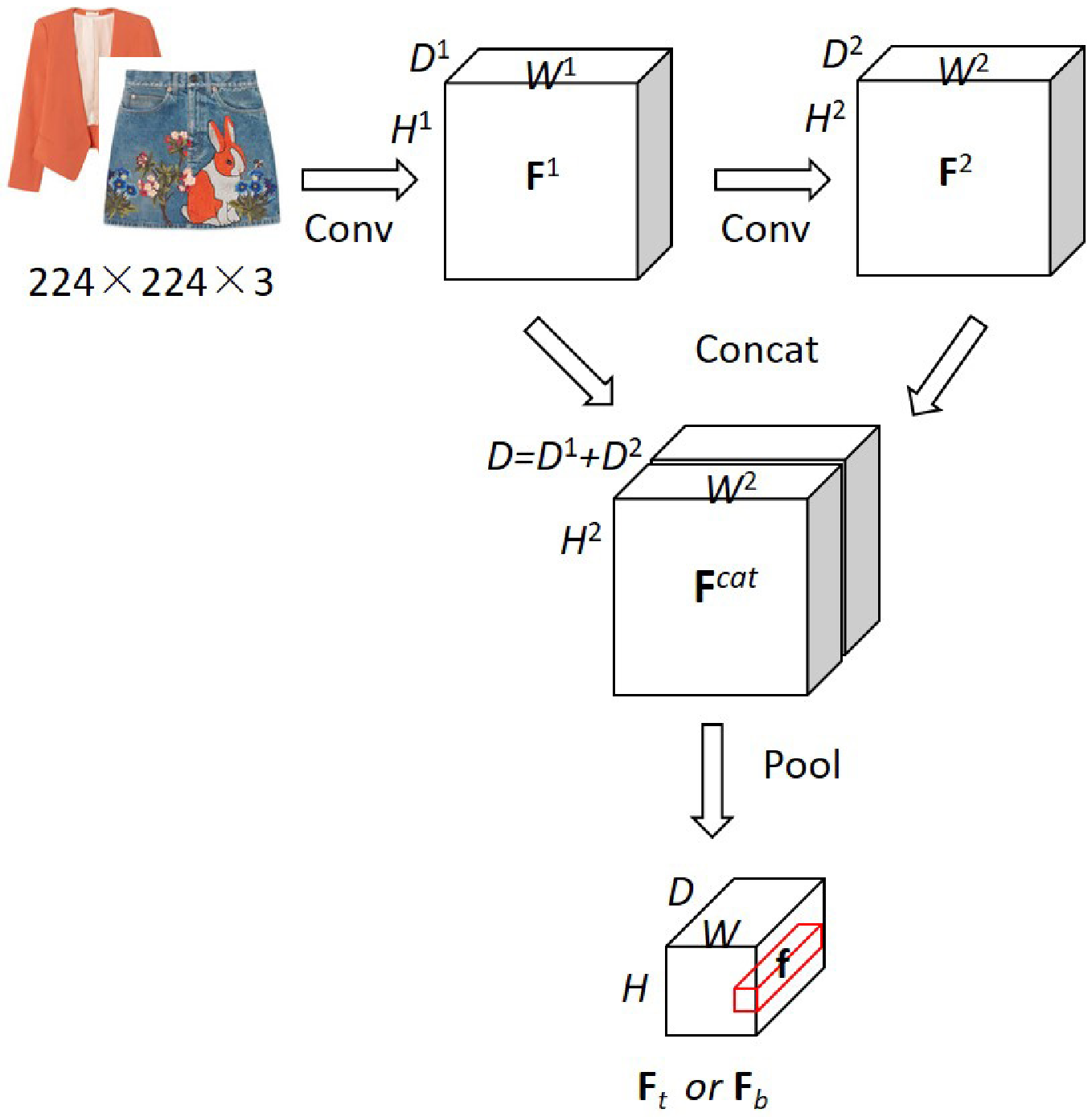}}
 \qquad\qquad
 \subfigure[Mutual attention and matching decoder.]{
 \label{f_3_2b}
 \includegraphics[width=0.2\textwidth]{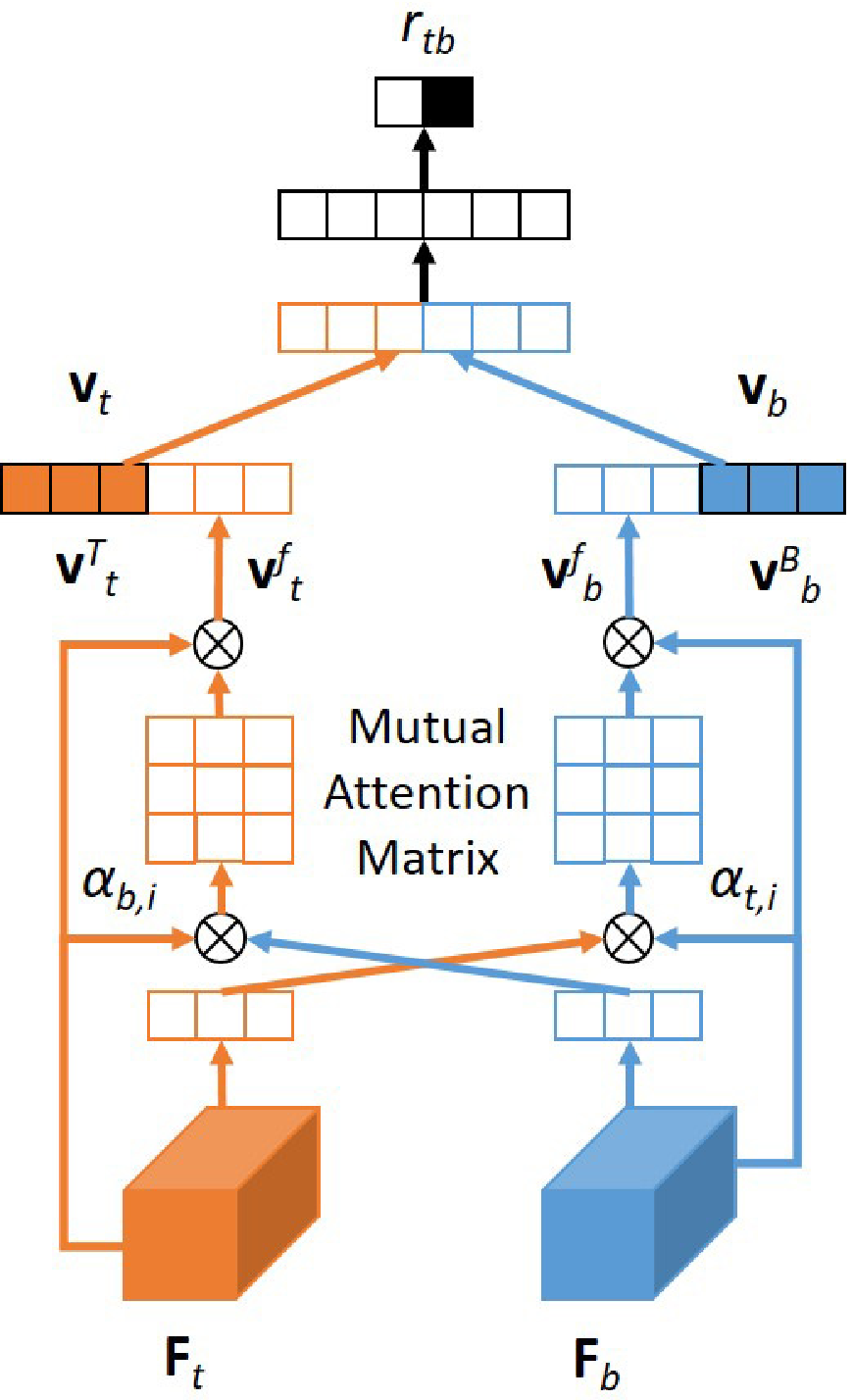}}
 \qquad\qquad
  \subfigure[Cross-modality attention and generation decoder.]{
  \label{f_3_2c}
 \includegraphics[width=0.3\textwidth]{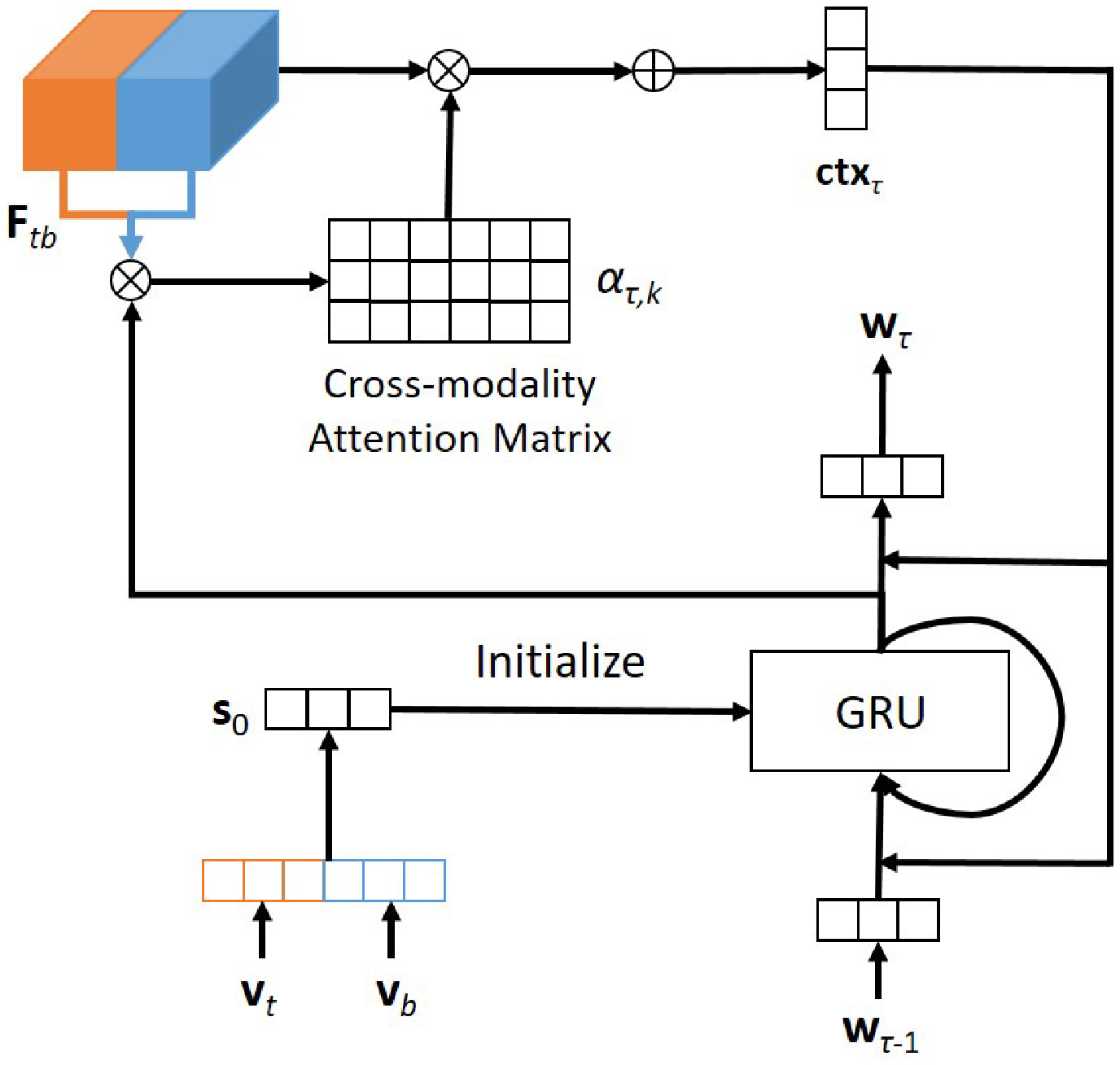}}
 \caption{Details of the neural fashion recommendation architecture (\ac{NOR}).
(a)~The top and bottom image encoder extracts visual features $\textbf{F}_t$ and $\textbf{F}_b$ from images. (b)~Using the mutual attention mechanism, we transform visual features to  latent representations $\textbf{v}_t$ and $\textbf{v}_b$. Then the matching decoder predicts the matching indicator $r_{tb}$. (c)~At each timestamp $\tau$, the generation decoder employs a cross-modality attention mechanism to generate the word $\textbf{w}_{\tau}$.}
 \label{f_3_2}
\end{figure*}

\subsection{Overview}
Given a top $t_i$ from a pool $\mathcal{
T} = \{t_1, t_2, \ldots , t_{N_t}\}$, the \emph{bottom item recommendation task} is to recommend a ranked list of bottoms from a candidate pool $\mathcal{B} = \{b_1, b_2, \ldots , b_{N_b}\}$.
Similarly, the \emph{top item recommendation task} is to recommend a ranked list of tops for a given bottom.
The \emph{comment generation task} is to generate a natural-sounding comment $c^\mathit{tb}$ for each recommended outfit (i.e., top-bottom pair).
The generated comments can be regarded as explanations for each recommended outfit: why is an outfit  matched? 
Note that it does not matter whether we perform bottom item recommendation or top item recommendation, \ac{NOR} generates similar comments for the same outfit, because the generated comments are for the whole outfit.

As shown in Fig.~\ref{f_3_1}, \ac{NOR} consists of three core components, a \emph{top and bottom image encoder}, a \emph{matching decoder}, and a \emph{generation decoder}.
Based on a convolutional neural network~\cite{Lecun1998Gradient}, the top and bottom image encoder (Fig.~\ref{f_3_2a}) extracts visual features from images including a pair $(t,b)$, and transforms visual features to the latent representations of $t$ and $b$, respectively.
A mutual attention mechanism is introduced here to guarantee that the top and bottom image encoder can encode the compatibility between $t$ and $b$ into their latent representations.
In Fig.~\ref{f_3_2b}, the matching decoder is a \ac{MLP} that evaluates the matching score between $t$ and $b$.
The generation decoder in Fig.~\ref{f_3_2c} is a \ac{GRU} \cite{Cho2014On}, which is used to translate the combination of the latent representation of a top and the latent representation of a bottom into a sequence of words as comments.
For the generation decoder, we propose cross-modality attention to better model the transformation between the visual and textual space.

Next, we detail each of the three core components.

\subsection{Top and bottom image encoder}
The top image encoder and the bottom image encoder are \acp{CNN}, which have been widely used in outfit recommendation~\cite{McAuley2015,Song2018Neural}.
Although there are many powerful architectures, like ResNet~\cite{He2016Deep} or DenseNet~\cite{huang2017densely}, training these architectures is not easy, because they have many parameters and need a lot of data and time to train. 
To balance the training cost and the test performance, we design a two-layer \acp{CNN} with mutual attention as the top and bottom image encoder, which has fewer parameters and yields good performance.
We verify the effectiveness of our network architecture in experiments.

Given a pair of images $(I_t,I_b)$, we assume that image $I_t$ and image $I_b$ are of size $224\times{224}$ with $3$ channels.
As shown in Fig.~\ref{f_3_2a}, we extract visual features from $I_t$ or $I_b$ via a two-layer \ac{CNN}.
Specifically, we first feed $I_t$ or $I_b$ to a convolutional layer to get primary visual features $\textbf{F}^1\in\mathbb{R}^{H^1\times{W^1}\times{D^1}}$.
Then we feed $\textbf{F}^1$ into another convolutional layer to obtain advanced visual features $\textbf{F}^2\in\mathbb{R}^{H^2\times{W^2}\times{D^2}}$.
Given the lessons learned with DensetNet~\cite{huang2017densely} for utilizing visual features more efficiently in different \ac{CNN} layers, we make sure $H^1=H^2$ and $W^1=W^2$ with padding operations so that we can concatenate $\textbf{F}^1$ and $\textbf{F}^2$ to get $\textbf{F}^\mathit{cat}\in\mathbb{R}^{H^2\times{W^2}\times{(D^1+D^2)}}$.
Finally, we use max-pooling in $\textbf{F}^\mathit{cat}$ to obtain the final visual features $\textbf{F}\in\mathbb{R}^{H\times{W}\times{D}}$.

Then we reshape $\textbf{F} = [\textbf{f}^1, \ldots , \textbf{f}^L]$ by flattening the width and height of the original $\textbf{F}$, where $\textbf{f}^i\in\mathbb{R}^D$ and $L = W\times{H}$.
We can consider $\textbf{f}^i$ as the visual features of the $i$-th local region of the input image.
Given a pair consisting of a top image $I_t$ and bottom image $I_b$, they will be fed into the same \ac{CNN}, i.e., the top and bottom image encoder have the same structure and share parameters.
For $I_t$, the extracted visual features $\textbf{F}_t$ are denoted as in Eq.~\ref{F_t}:
\begin{equation}
\label{F_t}
\textbf{F}_t = [\textbf{f}^1_t, \ldots , \textbf{f}^L_t], \quad \textbf{f}^i_t\in\mathbb{R}^D.
\end{equation}%
Similarly, for the extracted visual features $\textbf{F}_b$ of image $I_b$,  we have:
\begin{equation}
\label{F_b}
\textbf{F}_b = [\textbf{f}^1_b, \ldots , \textbf{f}^L_b], \quad \textbf{f}^i_b\in\mathbb{R}^D.
\end{equation}%
Previous attention mechanisms~\cite{Bahdanau2014Neural,Luong2015Effective} are not specifically designed for outfit recommendation, so they are not suitable to model the mutual matching relation between top images and bottom images.
We propose mutual attention mechanism to evaluate the correlation and alignment between each local region of $I_t$ and $I_b$, as shown in Fig.~\ref{f_3_2b}.
Because mutual attention can model the matching relation from two sides, i.e. from bottom images to top images and from top images to bottom images. So, it is more suitable for outfit recommendation.
To calculate the attention weights of top to bottom, we first perform global-average-pooling in $\textbf{F}_t$, which aggregates the visual features from all local regions to get global visual features $\textbf{g}_t\in\mathbb{R}^D$ of $I_t$ in Eq.~\ref{g_t}:
\begin{equation}
\label{g_t}
\textbf{g}_t = \frac{1}{L}\sum^L_{i=1}\textbf{f}^i_t.
\end{equation}%
Then, for the $i$-th local region of $I_b$, we can calculate the attention weight $e_{t,i}$ with $\textbf{g}_t$ and $\textbf{f}^i_b$ as in Eq.~\ref{e_t_i} by following \cite{Bahdanau2014Neural}:
\begin{equation}
\label{e_t_i}
e_{t,i} = \textbf{v}^{\top}_a\tanh(\textbf{W}_a\textbf{f}^{i}_b+\textbf{U}_a\textbf{g}_t),
\end{equation}%
where $\textbf{W}_a$ and $\textbf{U}_a\in\mathbb{R}^{D\times{D}}$ and $\textbf{v}_a\in\mathbb{R}^{D}$.
The attention weights are normalized in Eq.~\ref{a_t_i}:
\begin{equation}
\label{a_t_i}
\alpha_{t,i} = \frac{\mathrm{exp}(e_{t,i})}{\sum^{L}_{i=1}\mathrm{exp}(e_{t,i})}.
\end{equation}
Then we calculate the weighted sum of $\textbf{f}^{i}_b$ by $\alpha_{t,i}$ to get the attentive global visual features $\textbf{g}^a_b\in\mathbb{R}^D$ of $I_b$:
\begin{equation}
\textbf{g}^a_b = \sum^L_{i=1}\alpha_{t,i}\textbf{f}^{i}_b.
\end{equation}%
Similarly, we can calculate the attention weights of bottom to top and obtain the attentive global visual features $\textbf{g}^a_t$ of $I_t$:
\begin{equation}
\begin{split}
\textbf{g}_b & = \frac{1}{L}\sum^L_{i=1}\textbf{f}^i_b,\quad
e_{b,i} = \textbf{v}^{\top}_a\tanh(\textbf{W}_a\textbf{f}^{i}_t+\textbf{U}_a\textbf{g}_b),\\
\alpha_{b,i} & = \frac{\mathrm{exp}(e_{b,i})}{\sum^{L}_{i=1}\mathrm{exp}(e_{b,i})},\quad
\textbf{g}^a_t = \sum^L_{i=1}\alpha_{b,i}\textbf{f}^{i}_t.
\end{split}
\end{equation}%
We then project $\textbf{g}^a_t$ and $\textbf{g}^a_b$ to visual feature vectors $\textbf{v}^f_t$ and $\textbf{v}^f_b\in\mathbb{R}^{m_v}$:
\begin{equation}
\begin{split}
&\textbf{v}^f_t = \mathrm{ReLU}(\textbf{W}_p\textbf{g}^a_t),\quad
\textbf{v}^f_b = \mathrm{ReLU}(\textbf{W}_p\textbf{g}^a_b),
\end{split}
\end{equation}%
where $\textbf{W}_p\in\mathbb{R}^{m_v\times{D}}$ and $m_v$ is the size of $\textbf{v}^f_t$ and $\textbf{v}^f_b$.

Finally, building on insights from matrix factorization-based methods \cite{Koren2009Matrix,Lee2000Algorithms,Salakhutdinov2007Probabilistic}, we also learn top latent factors $\textbf{T}\in\mathbb{R}^{N_T\times{m_v}}$ and bottom latent factors $\textbf{B}\in\mathbb{R}^{N_b\times{m_v}}$ directly through which we incorporate collaborative filtering information as a complement to visual features.
Specifically, for each top $t$ and each bottom $b$, we have latent factors $\textbf{v}^T_t$ and $\textbf{v}^B_b$:
\begin{equation}
\label{T&B}
\textbf{v}^T_t = \textbf{T}(t,:), \quad \textbf{v}^B_b = \textbf{B}(b,:),
\end{equation}%
where $\textbf{v}^T_t$ and $\textbf{v}^B_b\in\mathbb{R}^{m_v}$. And we concatenate visual feature vectors and latent factors to get the latent representations $\textbf{v}_t$ and $\textbf{v}_b$:
\begin{equation}
\label{eqlar}
\begin{split}
\textbf{v}_t = [\textbf{v}^f_t,\textbf{v}^T_t],\quad \textbf{v}_b = [\textbf{v}^f_b,\textbf{v}^B_b],
\end{split}
\end{equation}%
where $\textbf{v}_t$ and $\textbf{v}_b\in\mathbb{R}^m$, $m=2m_v$.

\subsection{Matching decoder}
As shown in Fig.~\ref{f_3_2b}, we employ a multi-layer neural network to calculate the matching probability of $t$ and $b$. Given latent representations $\textbf{v}_t$ and $\textbf{v}_b$ calculated in Eq.~\ref{eqlar}, we first map $\textbf{v}_t$ and $\textbf{v}_b$ into a shared space:
\begin{equation}
\label{h_r}
\textbf{h}_r = \mathrm{ReLU}(\textbf{W}_s\textbf{v}_t+\textbf{U}_s\textbf{v}_b),
\end{equation}%
where $\textbf{h}_r\in\mathbb{R}^n$, and $\textbf{W}_s$ and $\textbf{U}_s\in\mathbb{R}^{n\times{m}}$ are the mapping matrices for $\textbf{v}_t$ and $\textbf{v}_b$, respectively.
Then we estimate the matching probability as follows:
\begin{equation}
\label{p_r_tb}
p(r_{tb}) = \mathrm{softmax}(\textbf{W}_r\textbf{h}_r),
\end{equation}%
where $\textbf{W}_r\in\mathbb{R}^{2\times{n}}$, and $p(r_{tb})\in\mathbb{R}^{2}$ which provides the probability distribution in $r_{tb}=1$ (corresponding to $p(r_{tb}=1)$) and $r_{tb}=0$ (corresponding to $p(r_{tb}=0)$).
Here, $r_{tb}=1$ denotes that $t$ and $b$ match and $r_{tb}=0$ denotes that $t$ and $b$ do not match.
Finally, we can recommend tops or bottoms according to $p(r_{tb})$.

\subsection{Generation decoder}
Following existing studies~\cite{NiLipVikMcA17,Li2017}, we also use \acp{RNN} to generate comments.
As shown in Fig.~\ref{f_3_2c}, we employ a \ac{GRU} with cross-modality attention as the generation decoder.
First, we compute the initial hidden state $\textbf{s}_0$ for the generation decoder with $\textbf{v}_t$ and $\textbf{v}_b$ in Eq.~\ref{s_0}:
\begin{equation}
\label{s_0}
\textbf{s}_0 = \tanh(\textbf{W}_i\textbf{v}_t+\textbf{U}_i\textbf{v}_b),
\end{equation}%
where $\textbf{s}_0\in\mathbb{R}^q$, $\textbf{W}_i$ and $\textbf{U}_i\in\mathbb{R}^{q\times{m}}$, and $q$ is the hidden size of the GRU.
Then, at each time stamp $\tau$, the GRU reads the previous word embedding $\textbf{w}_{\mathsf{\tau-1}}$, the previous context vector $\textbf{ctx}_{\tau-1}$ and the previous hidden state $\textbf{s}_{\tau-1}$ as input to compute the new hidden state $\textbf{s}_{\tau}$ and the current output $\textbf{o}_{\tau}$ in Eq.~\ref{new_s}:
\begin{equation}
\label{new_s}
\textbf{s}_{\tau},\textbf{o}_{\tau} = \mathrm{GRU}(\textbf{w}_{\tau-1},\textbf{ctx}_{\tau-1},\textbf{s}_{\tau-1}),
\end{equation}%
where $\textbf{w}_{\tau-1}\in\mathbb{R}^e$, $\textbf{ctx}_{\tau-1}\in\mathbb{R}^D$, $\textbf{s}_{\tau}$ and $\textbf{o}_{\tau}\in\mathbb{R}^q$, and $e$ is the word embedding size.
The context vector $\textbf{ctx}_{\tau}$ for the current timestamp $\tau$ is the weighted sum of all visual features from $\textbf{F}_t$ and $\textbf{F}_b$ and computed through the cross-modality attention.
It matches the current state $\textbf{s}_{\tau}$ with each element of $\textbf{F}_t$ and $\textbf{F}_b$ to get an importance score which makes better use of the extracted visual features to generate comments by paying attention to particularly effective visual features.
Recall that $\textbf{F}_t = [\textbf{f}^1_t, \ldots , \textbf{f}^L_t]$ and $\textbf{F}_b = [\textbf{f}^1_b, \ldots , \textbf{f}^L_b]$; we put them together as follows:
\begin{equation}
\textbf{F}_{tb} = [\textbf{f}^1_{tb}, \ldots , \textbf{f}^{2L}_{tb}], \quad \textbf{f}^{i}_{tb}\in\mathbb{R}^D.
\end{equation}%
The context vector $\textbf{ctx}_{\tau}$ is then computed by following \cite{Luong2015Effective}:
\begin{equation}
\begin{split}
&e_{\tau,k} = \textbf{s}^{\top}_{\tau}\textbf{W}_g\textbf{f}^{k}_{tb},\quad
\alpha_{\tau,k} = \frac{\exp(e_{\tau,k})}{\sum^{2L}_{k=1}\exp(e_{\tau,k})},\\
&\textbf{ctx}_{\tau} = \sum^{2L}_{k=1}\alpha_{\tau,k}\textbf{f}^{k}_{tb},
\end{split}
\end{equation}%
where $\textbf{W}_g\in\mathbb{R}^{q\times{D}}$.
Then $\textbf{o}_{\tau}$ and $\textbf{ctx}_{\tau}$ are used to predict the $\tau$-th word in Eq.~\ref{p_w}:
\begin{equation}
\label{p_w}
p(\textbf{w}_{\tau}|\textbf{w}_1,\textbf{w}_2,\ldots ,\textbf{w}_{\tau-1}) = \mathrm{softmax}(\textbf{W}_o\textbf{o}_{\tau}+\textbf{U}_o\textbf{ctx}_{\tau}),
\end{equation}%
where $\textbf{W}_o\in\mathbb{R}^{|V|\times{q}}$, $\textbf{U}_o\in\mathbb{R}^{|V|\times{D}}$, and $V$ is the vocabulary.

\subsection{Multi-task learning framework}
We use \ac{NIL} for both the matching task and generation task.
For the matching task, we define the loss function as follows:
\begin{equation}
L_\mathit{mat} = \sum_{\{r_{tb}\mid (t,b)\in{\mathcal{P}^+\cup\mathcal{P}^-}\}}-\log p(r_\mathit{tb}),
\end{equation}
where $\mathcal{P}^+ = \{(t_{i_1},b_{j_1}), (t_{i_2},b_{j_2}), \ldots , (t_{i_N},b_{j_N})\}$, $t_i\in\mathcal{T}$, $b_j\in\mathcal{B}$ is the set of positive combinations, which are top-bottom pairs extracted from the outfit combinations on Polyvore.
$\mathcal{P}^- = \{(t,b)\mid t\in\mathcal{T},b\in\mathcal{B}\wedge(t,b)\notin\mathcal{P}^+\}$ is the set of negative combinations, which are formed by tops and bottoms sampled randomly. 
Here, for positive combinations, $p(r_{tb})$ means the probability of $p(r_{tb}=1)$, i.e., the given pair matches; for negative pairs, $p(r_{tb})$ means the probability of $p(r_{tb}=0)$, i.e., the given pair does not match.

As for the generation task, the loss function is defined in Eq.~\ref{ssss}:
\begin{equation}
\label{ssss}
L_\mathit{gen} = \sum_{\{c^\mathit{tb}_k\mid c^\mathit{tb}_k\in{\mathcal{C}^\mathit{tb}}\wedge(t,b)\in\mathcal{P}^+\}}-\log p(c^\mathit{tb}_k),
\end{equation}
where $\mathcal{C}^\mathit{tb} = \{c^\mathit{tb}_1, c^\mathit{tb}_2, \ldots , c^\mathit{tb}_{N_{tb}}\}$ is the set of comments for each positive combinations of top $t$ and bottom $b$.
Note that we ignore the generation loss for negative combinations. We also add L2 loss as regularization to avoid overfitting:
\begin{equation}
L_\mathit{reg} = \|\Theta\|^2_2,
\end{equation}
where $\Theta$ is the set of neural parameters.
Finally, the multi-task objective function is a linear combination of $L_\mathit{mat}$, $L_\mathit{gen}$ and $L_\mathit{reg}$:
\begin{equation}
L = L_\mathit{mat}+L_\mathit{gen}+\lambda_\mathit{reg}L_\mathit{reg},
\end{equation}
where $\lambda_\mathit{reg}$ is used to adjust the weight of the regularization term. The whole framework can be efficiently trained using back-propagation in an end-to-end paradigm.


\section{Experimental Setup}
We set up experiments aimed at assessing the recommendation and generation performance; details shared between the two experiments are presented below.

\subsection{Datasets}
\begin{table}[]
\centering
\caption{Dataset statistics.}
\label{dataset}
\begin{tabular}{lrrrr}
\toprule
Dataset         & Tops   & Bottoms & Outfits & Comments \\ \midrule
WoW~\cite{Liu2012}               & 17,890 & 15,996  & 24,417  &     --    \\
Exact Street2Shop~\rlap{\cite{7410739}}~ &   --    &    --    & 39,479  &     --    \\
Fashion-136K~\cite{Jagadeesh2014}      &   --    &    --    & 135,893 &     --    \\
FashionVC~\cite{Song2017}         & 14,871 & 13,663  & 20,726  &     --    \\
\midrule
ExpFashion       & 29,113 & 20,902  & 200,745 & 1,052,821 \\
\bottomrule
\end{tabular}
\end{table}

In this section, we briefly introduce existing datasets and detail how we build our own dataset, \textit{ExpFashion}.

Existing fashion datasets include \textit{WoW} \cite{Liu2012}, \textit{Exact Street2Shop} \cite{7410739}, \textit{Fashion-136K} \cite{Jagadeesh2014}, and \textit{FashionVC} \cite{Song2017} datasets.
\textit{WoW}, \textit{Exact Street2Shop}, and \textit{Fashion-136K} are collected from street photos\footnote{\url{http://www.tamaraberg.com/street2shop/}} and thus inevitably involve a clothing parsing technique, which still remains a great challenge in the computer vision domain \cite{Song2017,Yamaguchi2012,6888484}.
Even though \textit{FashionVC} is crawled from Polyvore, it lacks user comments.
Moreover, the small scale of all existing datasets makes them insufficient for text generation.
We employ \textit{FashionVC} only to evaluate the recommendation part. 

To be able to evaluate the recommendation and generation results, we collected a large dataset from Polyvore.
In particular, starting from 1,000 seed outfits, we crawled new outfits given an item from existing outfits, and stored outfits in the dataset, iteratively.
To balance quality and quantity, we only considered outfits with comments longer than 3 words.
We also removed tops or bottoms with fewer than 3 occurrences.
We ended up with 200,745 outfits with 29,113 tops, 20,902 bottoms, and 1,052,821 comments.
We randomly selected 1,000 tops and bottoms as validation set, 2,000 tops and bottoms as test set, and the remainder as training set.
Most selected tops and bottoms with their positive bottoms and tops in validation and test set are unpopular , and are not seen in training set. 
Here ``unpopular'' means the positive combinations of a top or a bottom is less than 10.
Since it is time consuming to evaluate each top-bottom pair, we followed existing studies \cite{Song2017} and randomly selected bottoms to generate 100 candidates along with the positive bottoms for each top in validation and test set.
For each top in the validation or test set, the positive bottoms are those that have been matched with the top on Polyvore, which form our ground truth for recommendation; they should be more in line with fashion than other candidates.
The same is true for both bottom item recommendation and top item recommendation.
For the generation task, we use all actual user comments of each outfit in the \textit{ExpFashion} dataset as the references to evaluate the generated comments.
The statistics of \textit{ExpFashion} are listed in Table~\ref{dataset}; for comparison we also list the characteristics of datasets used in previous work.

We also harvested other domains of information, e.g., visual images, categories, title description, etc., and other kinds of items, e.g., shoes, accessories, etc. 
All this information can be employed for future research.\footnote{The dataset is available at \url{https://bitbucket.org/Jay_Ren/fashion_recommendation_tkde2018_code_dataset}}
Because there is no user information (ID or any other information) in the datasets, none of the models (the baselines and our own model) in both recommendation and generation are personalized models.
There are no labels in the ExpFashion dataset that identify whether a comment is for the top or bottom; most user comments are for the whole top-bottom pairs, so all models generate comments for complete outfits.

\subsection{Implementation details}
For the networks in the top and bottom image encoder, we set the kernel size of all convolutional layers to $3\times{3}$, the stride to 1, the padding to 1, the activation function to $\mathrm{relu}$, and the pooling size to $16\times{16}$. 
As a result, we have $H^1=H^2=W^1=W^2=224$, $D^1=D^2=32$, $H=W=14$ and $D=D^1+D^2=64$. 
The latent representation size $m$ is searched in $[200, 400, 600]$, but there is no significant difference; we set $m$ to 600.
For the matching decoder, we set the shared space size $n$ to 256.
The input and output vocabularies are collected from user comments, which have 92,295 words.
We set the word embedding size $e$ to 300 and all \ac{GRU} hidden state sizes $q$ to 512.
The regularization weight $\lambda_\mathit{reg}$ is searched in $[0.00001, 0.0001, 0.001, 0.01]$, where $0.0001$ is the best.
During training, we initialize model parameters randomly using the Xavier method \cite{Glorot2010Understanding}.
We use Adam \cite{Kingma2014Adam} as our optimization algorithm.
For the hyper-parameters of the Adam optimizer, we set the learning rate $\alpha$ = 0.001, two momentum parameters $\beta$1 = 0.9 and $\beta$2 = 0.999 respectively, and $\epsilon$ = $10^{-8}$.
We also apply gradient clipping \cite{Pascanu2013} with range $[-5, 5]$ during training.
To both speed up the training and converge quickly, we use mini-batch size 64 by grid search.
We test the model performance on the validation set for every epoch.
Because there is no negative outfit in the dataset, we randomly sample a top or bottom for each positive outfit.
For negative samples, we do not train the comment generation part.
During testing, for comment generation, we use beam search \cite{Koehn2004Pharaoh} to get better results.
To avoid favoring shorter outputs, we average the ranking score along the beam path by dividing it by the number of generated words.
To balance decoding speed and performance, we set the beam size to 3. 
Our framework is implemented in Tensorflow~\cite{Abadi2015TensorFlow}; the code is available at \url{https://bitbucket.org/Jay_Ren/fashion_recommendation_tkde2018_code_dataset}. 
All experiments were conducted on a single Titan X GPU.


\section{Bottom 
and Top Item Recommendation}
\label{section:recommendation}
In this section, we present our experimental results on the recommendation task.
We first specify the experimental details for this task. 
Then we discuss experimental results on the ExpFashion and FashionVC datasets, respectively.

\subsection{Methods used for comparison}
\label{subsection:recommendation_baselines}
We consider the following baselines in the bottom and top item recommendation experiments.

\begin{itemize}[nosep,leftmargin=*]
\item \textit{POP}: POP simply selects the most popular bottoms for each top and vice versa. Here, ``popularity'' is defined as the number of tops that have been paired with the bottom. POP is frequently used as a baseline in recommender systems \cite{He2017NCF}.

\item \textit{NRT}: NRT~\cite{Li2017} introduces recurrent neural networks into collaborative filtering. 
It can jointly predict ratings and generate tips based on latent factors of users and items. 
For comparison, we adapt NRT to make it compatible with outfit recommendation. 
The input of NRT are the IDs of a top and a bottom, and the output are the comments for this top-bottom pair, and the matching score between the given top and bottom rather than the rating. 
And the number of hidden layers for the regression part is set to 1. 
The beam size is set to 3. 
In addition, because there are no reviews in our datasets, we remove the relative part from NRT. 
Other configurations follow the original paper.
We do not only compare the recommendation performance of the models we consider, but also compare the quality of the generated comments, see Section~\ref{subsection:generation_baselines}.

\item \textit{DVBPR}: DVBPR~\cite{Kang2017Visually} employs the \ac{CNN}-F~\cite{Chatfield14} to learn image representations and jointly recommends fashion items to users by collaborative filtering. 
We modify DVBPR to make it work on our task. 
First we use \ac{CNN}-F to learn the image representations of the given top $t$ and bottom $b$.
Then we calculate the matching score $m_{tb}$ as follows:
\begin{equation}
m_{tb} = \textbf{v}_t^T\textbf{v}_b,
\end{equation}
where $\textbf{v}_t$ and $\textbf{v}_b$ are the image representations learned by \ac{CNN}-F, whose size are set to 100. 
Finally we also train DVBPR by \ac{BPR} loss.

\item \textit{SetRNN}:
SetRNN~\cite{Li2017MiningFO} trains AlexNet~\cite{Jia2014} to extract visual features from top images and bottom images.
And it adapts an \ac{RNN} as a pooling model to classify a given outfit as popular or unpopular based on the extracted features.
We change its target to predict whether a given outfit is matched or not.

\item \textit{IBR}: IBR~\cite{McAuley2015} models the relation between objects based on their visual appearance. This work also learns a visual style space, in which related objects are retrieved using nearest-neighbor search. 
In experiments, the embedding size of objects is set to 100.

\item \textit{BPR-DAE}: BPR-DAE~\cite{Song2017} is a content-based neural framework that models the compatibility between fashion items based on the Bayesian personalized ranking framework. BPR-DAE is able to jointly model the coherence relation between modalities of items and their implicit matching preference. 
We set the latent representation size of items to 512 in experiments.
\end{itemize}

\noindent%
Note that POP and NRT recommend items based on historical records to count the popularity and learn the latent factors respectively. 
So they cannot generalize well to new items that lack historical records \cite{Aharon2015ExcUseMe}. 
But DVBPR, SetRNN, IBR, BPR-DAE and \ac{NOR} model the matching relation between fashion items based on their image content and learn to recommend using visual features. 
As a result, they can generalize to new items as long as there are images.

\subsection{Evaluation metrics}
We employ three evaluation metrics in the bottom and top item recommendation experiments: \acfi{MAP}, \acfi{MRR}, and \acfi{AUC}.
All are widely used evaluation metrics in recommender systems~\cite{Rendle2010,Zhang2013,DBLP:conf/cikm/LiRCRLM17}.

As an example, in bottom item recommendation, MAP, MRR, and AUC are computed as follows,
\begin{equation}
\text{MAP}=\frac{1}{|T|}\sum_{i=1}^{|T|}\frac{1}{rel_i}\sum_{j=1}^{|B|}(P(j) \times rel(j)),
\end{equation}
where $B$ is the candidate bottom list; $P(j)$ is the precision at cut-off $j$ in the list; $rel_i$ is the number of all positive bottoms for top $i$; $rel(j)$ is an indicator function equaling 1 if the item at rank $j$ is a positive bottom, 0 otherwise.
\begin{equation}
\text{MRR}=\frac{1}{|T|}\sum_{i=1}^{|T|} \frac{1}{rank_i},
\end{equation}
where $rank_i$ refers to the rank position of the first positive bottom for the $i$-th top.
\begin{equation}
\mbox{}\hspace*{-2mm}
\text{AUC}=\frac{1}{|T|} \sum_{i=1}^{|T|} \frac{1}{|E(i)|}\sum_{(j,k)\in E(i)}\delta (f(t_i,b_j)> f(t_i,b_k)),
\end{equation}
where $T$ is the top collection as queries; $E(i)$ is the set of all positive and negative candidate bottoms for the $i$-th top; $\delta (\alpha)$ is an indicator function that equals 1 if $\alpha$ is true and 0 otherwise.

For significance testing we use a paired t-test with $p<0.05$.

\subsection{Results on the ExpFashion dataset}
\begin{table}[]
\centering
\caption{Results of bottom and top item recommendation on the ExpFashion dataset (\%).}
\label{rec_table1}
\begin{tabular}{l ccc ccc}
\toprule
\multirow{2}{*}{Method} & \multicolumn{3}{c}{Bottom item} & \multicolumn{3}{c}{Top item} \\
\cmidrule(r){2-4} \cmidrule{5-7}
& \multicolumn{1}{c}{MAP} & \multicolumn{1}{c}{MRR} & \multicolumn{1}{c}{AUC} & \multicolumn{1}{c}{MAP} & \multicolumn{1}{c}{MRR} & \multicolumn{1}{c}{AUC} \\
\midrule
POP & \phantom{1}5.45 & \phantom{1}6.45 & 49.19 & \phantom{1}6.91 & \phantom{1}9.16 & 51.71 \\
NRT & \phantom{1}6.36 & \phantom{1}8.54 & 49.49 & \phantom{1}7.74 & 11.78 & 50.98 \\
DVBPR & \phantom{1}8.55 & 12.10 & 57.96 & 11.08 & 16.98 & 60.31 \\
SetRNN & \phantom{1}6.60 & \phantom{1}8.86 & 51.69 & \phantom{1}7.07 & \phantom{1}9.93 & 51.89 \\
IBR & \phantom{1}7.30 & \phantom{1}9.99 & 52.60 & \phantom{1}8.22 & 12.54 & 52.39 \\
BPR-DAE & 10.09 & 13.89 & 61.36 & 11.51 & 17.73 & 61.75 \\
\ac{NOR} & \textbf{11.54}\rlap{$^\dagger$} & \textbf{15.38}\rlap{$^\dagger$} & \textbf{64.75}\rlap{$^\dagger$} & \textbf{13.48}\rlap{$^\dagger$} & \textbf{20.83}\rlap{$^\dagger$} & \textbf{65.09}\rlap{$^\dagger$} \\
\bottomrule
\end{tabular}
\begin{minipage}{\columnwidth}
\small
The superscript $^\dagger$ indicates that \ac{NOR} significantly outperforms BPR-DAE.
\end{minipage}
\end{table}

\begin{table}[]
\centering
\caption{Results of bottom and top item recommendation on the FashionVC dataset (\%).}
\label{rec_table2}
\begin{tabular}{l ccc ccc}
\toprule
\multirow{2}{*}{Method} & \multicolumn{3}{c}{Bottom item} & \multicolumn{3}{c}{Top item} \\
\cmidrule(r){2-4} \cmidrule{5-7}
& \multicolumn{1}{c}{MAP} & \multicolumn{1}{c}{MRR} & \multicolumn{1}{c}{AUC} & \multicolumn{1}{c}{MAP} & \multicolumn{1}{c}{MRR} & \multicolumn{1}{c}{AUC} \\
\midrule
POP & 4.61 & 5.50 & 30.10 & 3.83 & \phantom{1}4.62 & 27.13 \\
DVBPR & 7.99 & 8.82 & 57.46 & 7.83 & \phantom{1}8.66 & 57.32 \\
SetRNN & 6.04 & 6.45 & 51.75 & 5.66 & \phantom{1}6.27 & 52.34 \\
IBR & 6.29 & 6.74 & 53.98 & 6.68 & \phantom{1}7.38 & 52.61 \\
BPR-DAE & 8.44 & \textbf{9.34} & 60.62 & 8.03 & \phantom{1}8.95 & 60.05 \\
\ac{NOR}-CG & \textbf{8.50} & 9.12 & \textbf{64.17}\rlap{$^\dagger$} & \textbf{9.40} & \textbf{10.24} & \textbf{65.28}\rlap{$^\dagger$} \\
\bottomrule
\end{tabular}
\begin{minipage}{\columnwidth}
\small
The superscript $^\dagger$ indicates that \ac{NOR}-CG significantly outperforms BPR-DAE.
\end{minipage}
\end{table}

The outfit recommendation results of \ac{NOR} and the competing models on the ExpFashion dataset are given in Table~\ref{rec_table1}.
\ac{NOR} consistently outperforms all baseline methods in terms of MAP, MRR, and AUC metrics on the ExpFashion dataset.
From the results in the table, we have five main observations.
First, \ac{NOR} significantly outperforms all baselines; \ac{NOR} achieves the best result on all metrics. 
Although IBR and BPR-DAE employ pre-trained \acp{CNN} (both use AlexNet~\cite{Jia2014} trained on ImageNet\footnote{http://www.image-net.org/}) to extract visual features from images, they do not fine-tune the \acp{CNN} during experiments. 
However, we use \acp{CNN} as a part of our model, namely the top and bottom image encoder, and jointly train them with the matching decoder and generation decoder on the dataset.
We believe that this enables us to extract more targeted visual features from images for our task.
DVBPR and SetRNN also jointly train \acp{CNN} to extract visual features.
But for \ac{NOR}, we incorporate the mutual attention mechanism that explicitly models the compatibility between a top and a bottom; this mechanism allows us to learn more effective latent representations for tops and bottoms; see Section~\ref{section:giraffe} for a further analysis.
Moreover, \ac{NOR} can utilize the information of user comments to improve the performance of outfit recommendation.
In fact, visual features and user comments are two modalities to explain why a top and a bottom match.
\ac{NOR} captures this information with its multi-task learning model.
This multi-task learning setup makes recommendations more accurate; see Section~\ref{section:chameleon} for a further analysis.

Second, IBR and BPR-DAE both use a pre-trained \ac{CNN} to extract visual features as input, but BPR-DAE performs better. 
IBR only executes a linear transformation, while BPR-DAE uses a more sophisticated compatibility space using an autoencoder neural network.


Third, NRT does not perform well on most metrics. 
One important reason is that our dataset is very sparse, where a top or a bottom only has about 8 positive combinations on average. 
Under such conditions, NRT, which relies on collaborative filtering, cannot learn effective latent factors \cite{Qian2014Personalized,Wang2014Improving}.

Fourth, the performance of POP is the worst; the reason is that popularity cannot be used to reflect why a top and a bottom are matched.
In outfit recommendation, the visual feature plays a more decisive role.
Incorporating visual signals directly into the recommendation objective can make recommendation more accurate \cite{Kang2017Visually}.
Because they all use \acp{CNN} to extract visual features, DVBPR, SetRNN, IBR, BRP-DAE, and \ac{NOR} all outperform POP and NRT.

Fifth, all methods' top item recommendations are better than their bottom item recommendations.
This is because in our dataset the average number of positive tops that each bottom has is larger than the average number of positive bottoms that each top has.
This makes bottom item recommendation more difficult than top item recommendation.

\subsection{Results on the FashionVC dataset}
\label{subsection:FashionVC}
\begin{table*}[h]
\centering
\caption{Results on the comment generation task (\%).}
\label{gen_table1}
\begin{tabular}{l ccc ccc ccc ccc c}
\toprule
\multirow{2}{*}{Methods} & \multicolumn{3}{c}{ROUGE-1} & \multicolumn{3}{c}{ROUGE-2} & \multicolumn{3}{c}{ROUGE-L} & \multicolumn{3}{c}{ROUGE-SU4} & \multirow{2}{*}{BLEU} \\
\cmidrule(r){2-4} \cmidrule(r){5-7} \cmidrule(r){8-10} \cmidrule{11-13}
& P & R & F1 & P & R & F1 & P & R & F1 & P & R & F1 & \\ \hline
LexRank & \textbf{9.60} & \phantom{1}8.88 & 8.17 & \textbf{2.51} & 2.23 & 2.09 & \textbf{9.12} & \phantom{1}8.38 & 7.73 & \textbf{4.43} & 3.65 & 3.05 & 30.55 \\
CTR & 7.16 & 11.43 & 7.95 & 2.01 & 2.91 & 2.17 & 6.69 & 10.57 & 7.39 & 2.95 & 5.22 & 3.10 & 27.43 \\
RMR & 7.46 & \textbf{12.26} & 8.44 & 2.02 & \textbf{3.00} & \textbf{2.23} & 6.91 & \textbf{11.27} & 7.78 & 2.95 & \textbf{5.49} & 3.22 & 28.46 \\
NRT & 7.75 & \phantom{1}8.98 & 7.71 & 1.80 &  2.30 & 1.83 & 7.52 & \phantom{1}8.74 & 7.48 & 3.05 & 3.93 & 2.78 & 35.61 \\
\ac{NOR} & 9.40\rlap{$^\dagger$} & $10.29^\dagger$ & \textbf{9.09}\rlap{$^\dagger$} & 2.21 & 2.27 & 2.05\rlap{$^\dagger$} & 8.85\rlap{$^\dagger$} & \phantom{1}9.68\rlap{$^\dagger$} & \textbf{8.55}\rlap{$^\dagger$} & 3.96\rlap{$^\dagger$} & 4.26\rlap{$^\dagger$} & \textbf{3.33}\rlap{$^\dagger$} & \textbf{37.21}\rlap{$^\dagger$} \\
\bottomrule
\end{tabular}
\begin{minipage}{\textwidth}
\small
The superscript $^\dagger$ indicates that our model \ac{NOR} performs significantly better than NRT as given by the 95\% confidence interval in the official ROUGE script.
\end{minipage}
\end{table*}

In order to confirm the effectiveness of our recommendation part, we also compare \ac{NOR}-CG, which is \ac{NOR} without the comment generation part, with POP, DVBPR, SetRNN, IBR and BPR-DAE on the FashionVC dataset; see Table~\ref{rec_table2}. Because there are no comments on FashionVC, we leave out NRT.

From Table~\ref{rec_table2}, we can see that \ac{NOR}-CG achieves the best performance in terms of the MAP and AUC scores on the bottom item recommendation task and also in terms of the MAP, MRR and AUC score on the top item recommendation task. 
\ac{NOR} is only slightly inferior to BPR-DAE in terms of MRR on the bottom item recommendation. 
This means that, even without the generation component, \ac{NOR}-CG can still achieve a better performance than other methods. 
Our top and bottom image encoder with mutual attention can extract effective visual features for outfit recommendation.

Note that only the differences in terms of AUC are significant. 
The reason is that the size of the FashionVC dataset is small. 
Although \ac{NOR}-CG achieves a 1.37\% and 1.29\% increase in terms of MAP and MRR respectively, it is hard to pass the paired t-test with a small test size.
\section{Comment Generation}
\label{section:comment-generation}
In this section, we assess the performance of comment generation.

\subsection{Methods used for comparison}
\label{subsection:generation_baselines}

No existing work on outfit recommendation is able to generate abstractive comments.
In order to evaluate the performance of \ac{NOR} and conduct comparisons against meaningful baselines, we refine existing methods to make them capable of generating comments as follows.

\begin{itemize}[nosep,leftmargin=*]
\item \textit{LexRank}: LexRank~\cite{Erkan2004} is an extractive summarization method.
We first retrieve all comments from the training set as a sentence collection.
Thereafter, given a top and a bottom, we merge relevant sentence collections into a single document.
Finally, we employ LexRank to extract the most important sentence from the document as the comment for this top-bottom pair.
\item \textit{CTR}: CTR~\cite{Wang2011} has been proposed for scientific article recommendation; it solves a one-class collaborative filtering problem. CTR contains a topic model component and it can generate topics for each top and each bottom. For a given top or bottom, we first select the top-30 words from the topic with the highest probability. Then, the most similar sentence from the same sentence collection that is used for LexRank is extracted. For a given outfit of a top and a bottom, we choose the one with the highest degree of similarity from the two extracted sentences of the top and the bottom as the final comment.
\item \textit{RMR}: RMR~\cite{Ling2014} utilizes a topic modeling technique to model review texts and achieves significant improvements compared with other strong topic modeling based methods.
We modified RMR to extract comments in the same way as CTR.
\item \textit{NRT}: We use the same settings as described above in Section~\ref{subsection:recommendation_baselines}.
\end{itemize}
Note that we give an advantage to LexRank, CTR, and RMR, since there are no comments available for many cases both in the experimental environment and in practice.

\subsection{Evaluation metrics}

We use ROUGE \cite{Lin2004} as our evaluation metric with standard options\footnote{ROUGE-1.5.5.pl -n 4 -w 1.2 -m -2 4 -u -c 95 -r 1000 -f A -p 0.5 -t 0} for the evaluation of abstractive comment generation.
It is a classical evaluation metric in the field of text generation \cite{Li2017} and counts the number of overlapping units between the generated text and the ground truth written by users.
The ROUGE-N score is defined as follows:
\begin{equation}
ROUGE\text{-}N_{recall} =\sum_{g_n \in \tilde{c}}\frac{C_{co}(g_n)}{\sum_{g_n \in c}C(g_n)},
\end{equation}
\noindent where $\tilde{c}$ is the generated comment; $c$ is the ground truth comment; $g_n$ is an n-gram; $C(g_n)$ is the number of n-grams in $\tilde{c}$; $C_{co}(g_n)$ is the number of n-grams co-occurring in $\tilde{c}$ and $c$.
{$ROUGE\text{-}N_{precision}$} is computed by replacing $c$ with $\tilde{c}$ in {$ROUGE\text{-}N_{recall}$}.
{ROUGE-L calculates the longest common subsequence between the generated comment and the true comment. And Rouge-SU4 counts the skip-bigram plus unigram-based co-occurrence statistics.}
We use Recall, Precision, and F-measure of ROUGE-1, ROUGE-2, ROUGE-L, and ROUGE-SU4 to evaluate the quality of the generated comments.
We also use BLEU \cite{Papineni2002} as another evaluation metric, which is defined as follows:
\begin{equation}
BLEU=BP\cdot \exp\left(\sum_{n=1}^{N}w_n\log p_n\right),
\end{equation}
where $w_n$ is the weight of the $n$-th word; $p_n$ is n-gram precision, which is computed as {$ROUGE\text{-}N_{precision}$}; $BP$ is the brevity penalty:
\begin{equation}
BP=
\begin{cases}
1, & \text{if } c > r\\\
e^{1-r/c}, & \text{if } c \leq r,
\end{cases}
\end{equation}
where $c$ is the length of the generated text and $r$ is the length of the reference text.

\subsection{Results}
The evaluation results of our model and competing methods on the comment generation task on the ExpFashion dataset are given in Table~\ref{gen_table1}.
We report Recall, Precision, and F-measure (in percentage) of ROUGE-1, ROUGE-2, ROUGE-L, and ROUGE-SU4.
Additionally, we also report BLEU.

Based on the results reported in Table~\ref{gen_table1}, we have three main observations.
First, \ac{NOR} achieves good performance on the ExpFashion dataset.
Especially in terms of BLEU and F-measure of ROUGE-1, ROUGE-L and ROUGE-SU4, \ac{NOR} gets the best results.
\ac{NOR} is not the best performer on all metrics; for example, LexRank has better performance than \ac{NOR} in terms of ROUGE precision.
Also RMR's ROUGE recall is better than \ac{NOR}.
This is because LexRank prefers short sentences while RMR prefers long sentences.
In contrast, \ac{NOR} gets much better ROUGE F-measure and BLEU, which means \ac{NOR} can generate more appropriate comments.
In other words, \ac{NOR} achieves more solid overall performance than other models.
The reasons are two-fold.
On the one hand, \ac{NOR} has a top and bottom image encoder to encode information of visual features into the latent representations of tops and bottoms. So it makes the latent representations in \ac{NOR} more effective.
On the other hand, we employ a mutual attention mechanism to make sure that the generation decoder can better convert visual features into text to generate comments.

Second, one exception to the strong performance of \ac{NOR} described above is that \ac{NOR} performs relatively poorly in terms of ROUGE-2.
The possible reasons are: (1)~The user comments in our dataset are very short, only about 7 words in length on average.
Naturally, the model trained using this dataset cannot generate long sentences.
(2)~The mechanism of a typical beam search algorithm makes the model favor short sentences.
(3)~The extraction-based approaches favor the extraction of long sentences.
So with an increase in N in ROUGE-N, the performance of \ac{NOR} suffers and the superiority of extraction-based methods is clear.

Third, due to the sparsity of the dataset, NRT performs poorly on most metrics.

\begin{figure}[!t]
 \centering
 \subfigure[Bottom mutual attention.]{
 \includegraphics[clip,trim=0mm 0mm 0mm 0mm,width=0.23\textwidth]{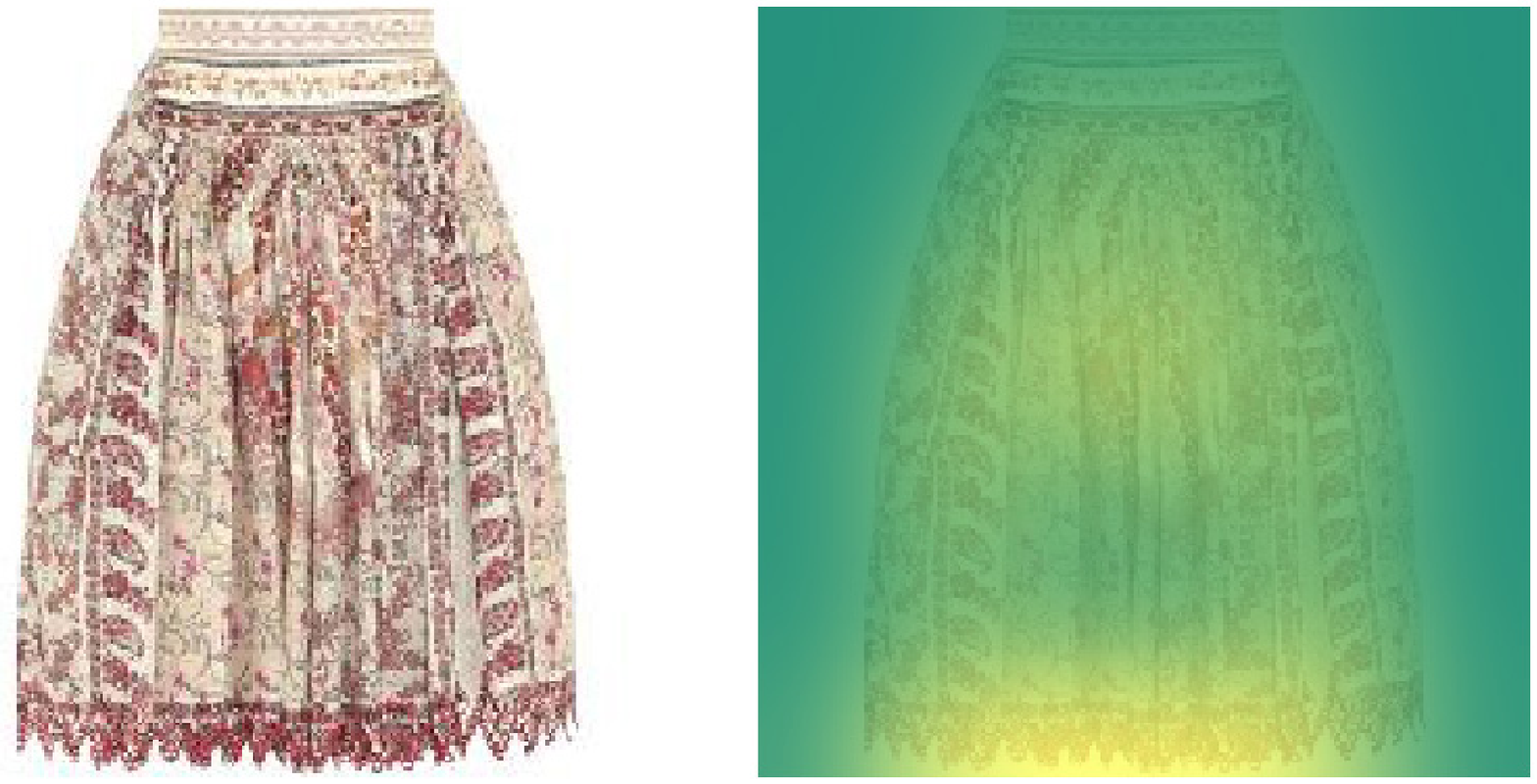}}
 \subfigure[Top mutual attention.]{
 \includegraphics[clip,trim=0mm 0mm 0mm 0mm,width=0.23\textwidth]{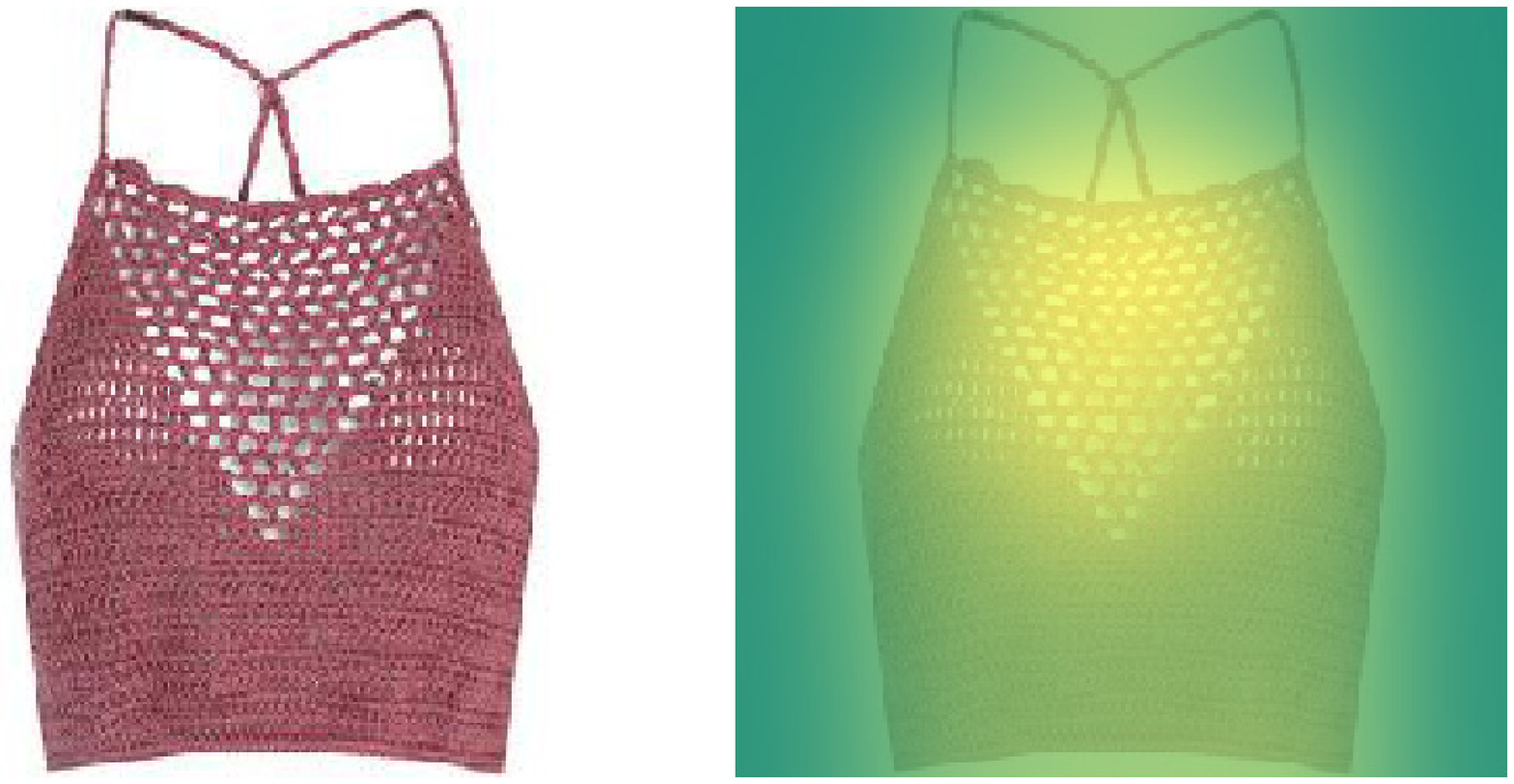}}
  \subfigure[Bottom cross-modality attention.]{
 \includegraphics[clip,trim=0mm 0mm 0mm 0mm,width=0.48\textwidth]{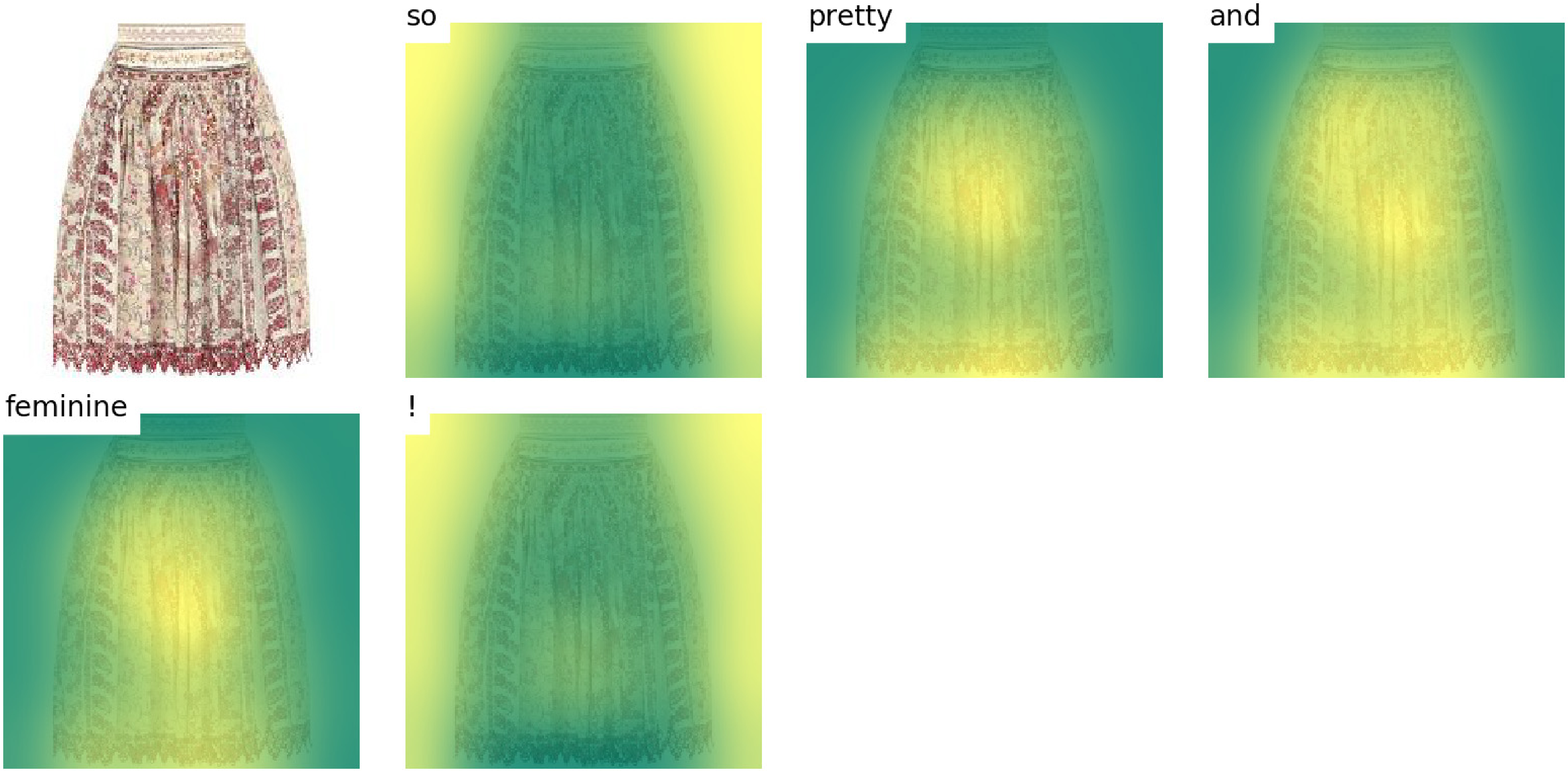}}
  \subfigure[Top cross-modality attention.]{
 \includegraphics[clip,trim=0mm 0mm 0mm 0mm,width=0.48\textwidth]{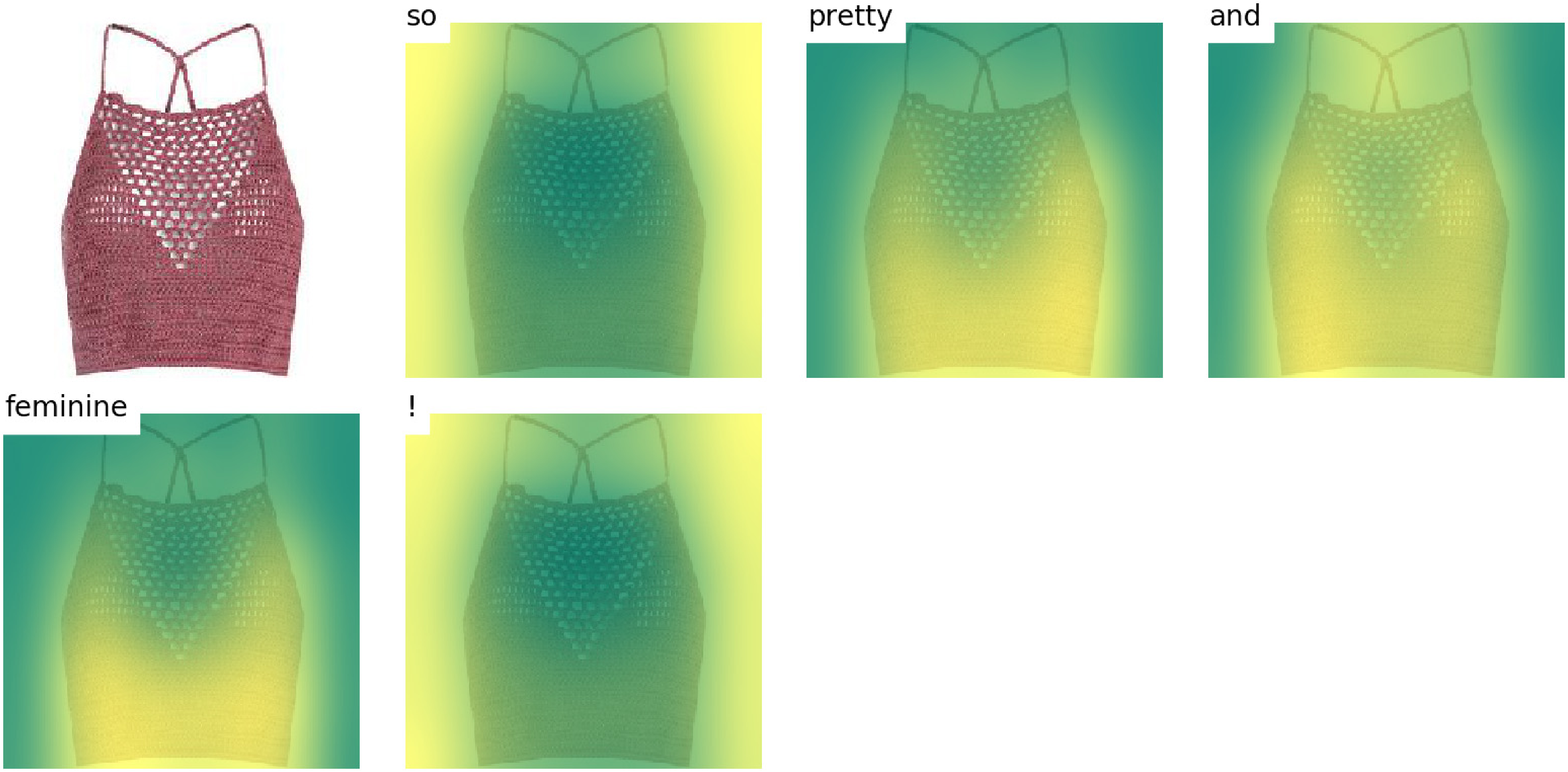}}
 \caption{Visualization of mutual attention and cross-modality attention.}
 \label{ana_figure1}
\end{figure}

\section{Analysis and Case Study}

In this section, we conduct further experiments to understand the effectiveness of attention, multi-task learning, and latent factors, followed by recommendation case studies and generation case studies.

\subsection{Attention mechanism analysis}
\label{section:giraffe}

To verify the effectiveness of the mutual attention mechanism and the cross-modality attention mechanism on the bottom and top item recommendation and comment generation tasks, we conduct experiments with different settings of \ac{NOR}.
The experimental results are shown in Table~\ref{ana_table1} and Table~\ref{ana_table2}.
\begin{table}[h]
\centering
\caption{Analysis of attention mechanisms on the bottom and top item recommendation tasks (\%).}
\label{ana_table1}
\begin{tabular}{l ccc ccc}
\toprule
\multirow{2}{*}{Attention} & \multicolumn{3}{c}{Bottom item} & \multicolumn{3}{c}{Top item} \\
\cmidrule(r){2-4} \cmidrule{5-7}
& MAP & MRR & AUC & MAP & MRR & AUC \\
\midrule
\ac{NOR}-NO & 10.96 & 14.93 & 64.72 & 13.38 & 20.13 & 65.40 \\
\ac{NOR}-MA & \textbf{12.55} & \textbf{16.98} & \textbf{67.13} & \textbf{14.65} & \textbf{21.58} & \textbf{66.98} \\
\ac{NOR}-CA & 11.72 & 15.48 & 64.85 & 13.53 & 20.71 & 65.60 \\
\ac{NOR}-(MA+CA) & 11.54 & 15.38 & 64.75 & 13.48 & 20.83 & 65.09 \\
\bottomrule
\end{tabular}
\begin{minipage}{\columnwidth}
\small
\end{minipage}
\end{table}
\begin{table*}[]
\centering
\caption{Analysis of attention mechanisms on comment generation (\%).}
\label{ana_table2}
\begin{tabular}{l ccc ccc ccc ccc c}
\toprule
\multirow{2}{*}{Attention} & \multicolumn{3}{c}{ROUGE-1} & \multicolumn{3}{c}{ROUGE-2} & \multicolumn{3}{c}{ROUGE-L} & \multicolumn{3}{c}{ROUGE-SU4} & \multirow{2}{*}{BLEU} \\
\cmidrule(r){2-4} \cmidrule(r){5-7} \cmidrule(r){8-10} \cmidrule{11-13}
& P & R & F1 & P & R & F1 & P & R & F1 & P & R & F1 & \\
\midrule
\ac{NOR}-NO & 8.17 & \phantom{1}9.14 & 7.99 & 2.03 & 2.45 & 2.00 & 7.84 & 8.83 & 7.69 & 3.32 & 4.21 & 2.99 & 34.07 \\
\ac{NOR}-MA & 8.00 & \phantom{1}9.36 & 8.00 & 2.10 & \textbf{2.58} & \textbf{2.09} & 7.67 & 9.04 & 7.69 & 3.27 & 4.40 & 3.03 & 32.75 \\
\ac{NOR}-CA & 8.42 & 10.08 & 8.54 & 2.12 & 2.56 & \textbf{2.09} & 7.97 & 9.59 & 8.09 & 3.42 & \textbf{4.59} & 3.18 & 34.37 \\
\ac{NOR}-(MA+CA) & \textbf{9.40} & \textbf{10.29} & \textbf{9.09} & \textbf{2.21} & 2.27 & 2.05 & \textbf{8.85} & \textbf{9.68} & \textbf{8.55} & \textbf{3.96} & 4.26 & \textbf{3.33} & \textbf{37.21} \\
\bottomrule
\end{tabular}
\end{table*}
From Table~\ref{ana_table1}, we notice that \ac{NOR}-MA (mutual attention only) performs better than \ac{NOR}-NO (no attention), not only on the bottom item recommendation task, but also on the top item recommendation task.
We conclude that the mutual attention mechanism can improve the performance of outfit recommendation.
Similarly, as shown in Table~\ref{ana_table2}, we observe that \ac{NOR}-CA (cross-modality attention only) outperforms \ac{NOR}-NO.
Thus we conclude that the cross-modality attention mechanism is helpful for the comment generation task.

In Table~\ref{ana_table1}, by comparing \ac{NOR}-MA with \ac{NOR}-(MA+CA), we also find that \ac{NOR}-MA outperforms \ac{NOR}-(MA+CA) on outfit recommendation.
That may be because the two kinds of attention mechanism can influence each other through joint training.
So in \ac{NOR}-(MA+CA) the mutual attention mechanism does not reach the same performance as \ac{NOR}-MA.
We think that this performance trade-off is worth making.
\ac{NOR}-(MA+CA) improves over \ac{NOR}-MA on comment generation in Table~\ref{ana_table2}.
Also, \ac{NOR}-(MA+CA) performs better than \ac{NOR}-NO on both outfit recommendation and comment generation in Table~\ref{ana_table1} and Table~\ref{ana_table2}, which means that the combination of the two attention mechanisms is effective.

We visualize the effects of both attention mechanisms~\cite{Xu2015Show}, as shown in Fig.~\ref{ana_figure1}. 
For bottom mutual attention, the hem of the skirt gets more attention. 
And for top mutual attention, \ac{NOR} pays more attention to the hollow grid on the vest. 
When generating comments, \ac{NOR} also knows which words are associated with fashion items. 
For example, when generating ``pretty and feminine'', both the top and the bottom get the main attention, because they are the description of the combination.
However, for ``so'' or ``!'' which are irrelevant to fashion items, \ac{NOR} pays little attention to the top and the bottom but to the background.
So by visualizing attention, we can see that \ac{NOR} knows how and when to use visual features of tops and bottoms to recommend items and generate comments.

\subsection{Multi-task learning analysis}
\label{section:chameleon}

To demonstrate that \ac{NOR} can use user comments to improve the quality of outfit recommendation by multi-task learning, we compare \ac{NOR} with \ac{NOR}-CG; see Table~\ref{ana_table3}.
\begin{table}[h]
\centering
\caption{Analysis of multi-task learning (\%).}
\label{ana_table3}
\begin{tabular}{l ccc ccc}
\toprule
\multirow{2}{*}{Methods} & \multicolumn{3}{c}{Bottom item} & \multicolumn{3}{c}{Top item} \\
\cmidrule(r){2-4} \cmidrule{5-7}
& \multicolumn{1}{c}{MAP} & \multicolumn{1}{c}{MRR} & \multicolumn{1}{c}{AUC} & \multicolumn{1}{c}{MAP} & \multicolumn{1}{c}{MRR} & \multicolumn{1}{c}{AUC} \\
\midrule
\ac{NOR}-CG  & 10.19 & 13.65 & 62.03 & 12.59 & 19.17 & 63.07 \\
\ac{NOR} & \textbf{11.54} & \textbf{15.38} & \textbf{64.75} & \textbf{13.48} & \textbf{20.83} & \textbf{65.09} \\
\bottomrule
\end{tabular}
\begin{minipage}{\columnwidth}
\end{minipage}
\end{table}
We can see that \ac{NOR} achieves significant improvements over \ac{NOR}-CG; on the bottom item recommendation task, MAP increases by 1.35\%, MRR increases by 1.73\%, AUC increases by 2.72\%, and on the top item recommendation task, MAP increases by 0.89\%, MRR increases by 1.66\%, AUC increases by 2.02\%.
Through joint learning, our multi-task framework \ac{NOR} learns shared representations \cite{Caruana1997} for both recommendation and generation, which can make effective use of the information in comments to improve recommendation performance.

Additionally, by comparing \ac{NOR}-CG in Table~\ref{ana_table3} with BPR-DAE in Table~\ref{rec_table1}, we find that, on ExpFashion, \ac{NOR}-CG also achieves comparable results to other methods, which is consistent with the results on FashionVC (see Section~\ref{subsection:FashionVC}).
On the bottom item recommendation task, \ac{NOR}-CG achieves 0.10\% and 0.67\% increases in MAP and AUC, respectively; and on the top item recommendation task, \ac{NOR}-CG achieves a 1.08\%, 1.44\% and 1.32\% increase in MAP, MRR and AUC, respectively. 
We conclude that the model structure of \ac{NOR} in the recommendation part is able to improve recommendation performance.

\subsection{Latent factors analysis}
\begin{table}[]
\centering
\caption{Analysis of latent factors (\%).}
\label{ana_table4}
\begin{tabular}{l ccc ccc}
\toprule
\multirow{2}{*}{Methods} & \multicolumn{3}{c}{Bottom item} & \multicolumn{3}{c}{Top item} \\
\cmidrule(r){2-4} \cmidrule{5-7}
& \multicolumn{1}{c}{MAP} & \multicolumn{1}{c}{MRR} & \multicolumn{1}{c}{AUC} & \multicolumn{1}{c}{MAP} & \multicolumn{1}{c}{MRR} & \multicolumn{1}{c}{AUC} \\
\midrule
\ac{NOR}-LF & \phantom{1}6.55 & \phantom{1}8.88 & 49.54 & \phantom{1}7.83 & 11.93 & 50.78 \\
\ac{NOR}-CG & 10.19 & 13.65 & 62.03 & 12.59 & 19.17 & 63.07 \\
\ac{NOR}-WLF & 10.62 & 14.69 & 62.17 & 12.28 & 18.73 & 62.69 \\
\ac{NOR} & \textbf{11.54} & \textbf{15.38} & \textbf{64.75} & \textbf{13.48} & \textbf{20.83} & \textbf{65.09} \\
\bottomrule
\end{tabular}
\begin{minipage}{\columnwidth}
\end{minipage}
\end{table}

To analyze the effect of latent factors $\textbf{T}$ and $\textbf{B}$ (see Eq.~\ref{T&B}) for recommendation, we compare \ac{NOR}-LF (\ac{NOR} that only uses item ID information) with \ac{NOR}-CG and \ac{NOR}.
As shown in Table~\ref{ana_table4}, we find that \ac{NOR}-LF does not perform very well. 
This is because it only uses latent factors to capture the information in the historical matching records and does not take content-based features into consideration.
Besides, due to the sparsity of the ExpFashion dataset, matrix factorization-based methods cannot work well.
By incorporating the visual information of images, \ac{NOR}-CG makes up for the deficiency of \ac{NOR}-LF and improves the recommendation performance.
Further, \ac{NOR} uses the textual information of comments to achieve better recommendations.

Meanwhile, we also compare \ac{NOR} with \ac{NOR}-WLF which is \ac{NOR} without the latent factors. 
In Table~\ref{ana_table4}, it shows that if there are no the latent factors, the MAP, MRR and AUC of \ac{NOR} all descend.
So we can draw a conclusion that the latent factors can capture the complementary information for the visual features to improve the recommendation performance.

\subsection{Recommendation case studies}
\begin{figure*}[]
 \centering
 \subfigure[Illustration of the bottom item recommendation.]{
 \label{f_10_1}
 \includegraphics[width=1.0\textwidth]{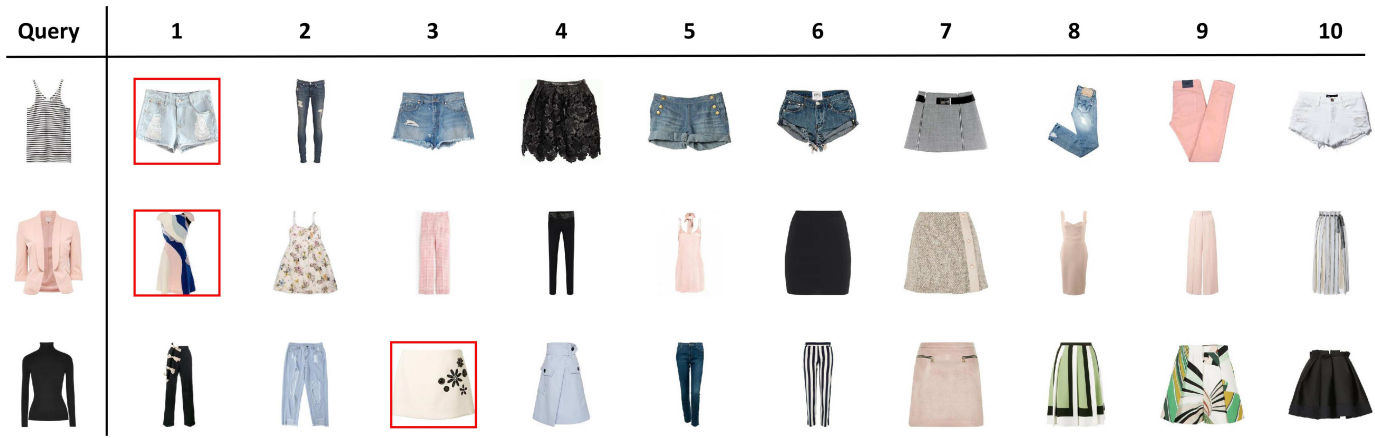}}
 \subfigure[Illustration of the top item recommendation.]{
 \label{f_10_2}
 \includegraphics[width=1.0\textwidth]{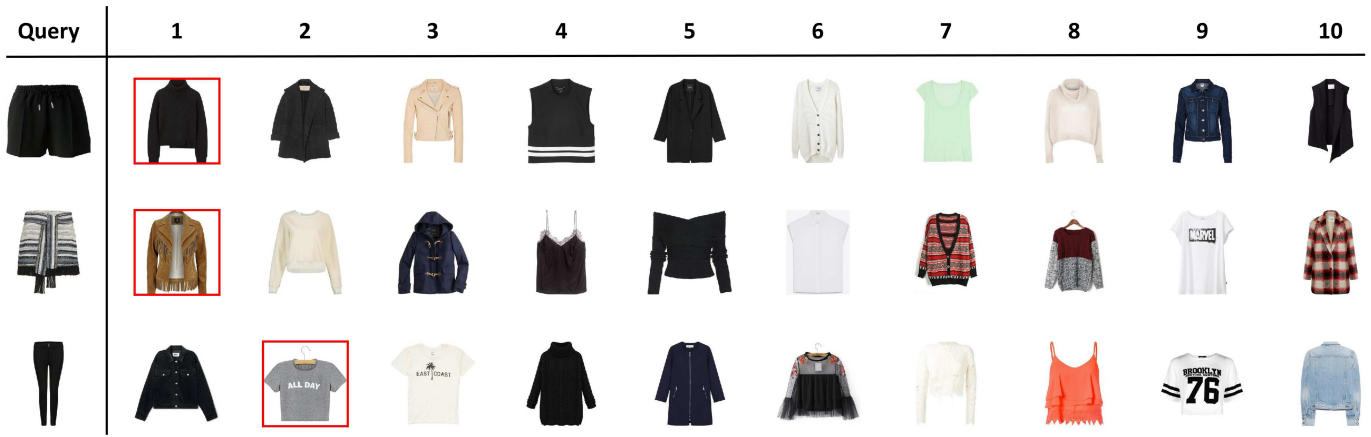}}
 \caption{Illustration of the recommendation results. The items highlighted in the red boxes are the positive ones.}
 \label{ana_figure2}
\end{figure*}

In Fig.~\ref{ana_figure2} we list some recommendation results of \ac{NOR} on the test set of ExpFashion. 
For each query item, we select the top-10 recommended items. 
And we use red boxes to highlight the positive items. 
Note that even if a recommended item is not highlighted with a red box, it should not be considered negative. We can see that most recommended items are compatible with the query items.
For example, the first given top seems to like denim shorts because the positive bottom is a light-colored denim shorts. 
So the recommended bottoms have many denim shorts or jeans. And the recommended skirts are also reasonable. 
Because they are short skirts and have similar shape with denim shorts. 
We also notice that sometimes \ac{NOR} cannot accurately rank the positive item at the first place. 
But the recommended items ranked before the positive item are also well enough for the given item, which is reasonable in real applications. 
For instance, for the last given bottom, the first top looks suitable  not only in color but also in texture.
Through these examples, we can see that \ac{NOR} can indeed provide good recommendations.

\subsection{Generation case studies}
\begin{table*}[]
\centering
\caption{Examples of recommendations and generated comments.}
\label{ana_table5}
\begin{tabular}{cc|cc|cc|cc}
\hline\\[-1em]
\includegraphics[width=0.1\textwidth]{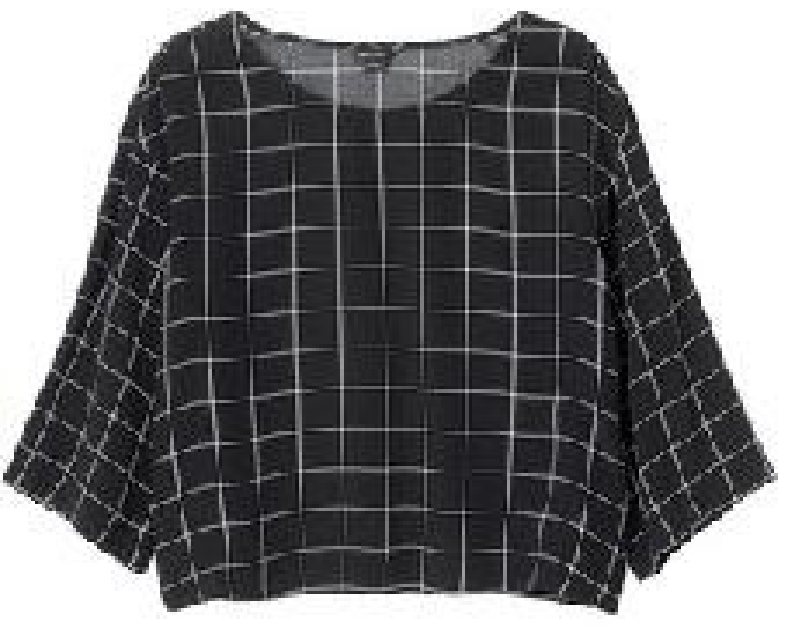} & \includegraphics[width=0.1\textwidth]{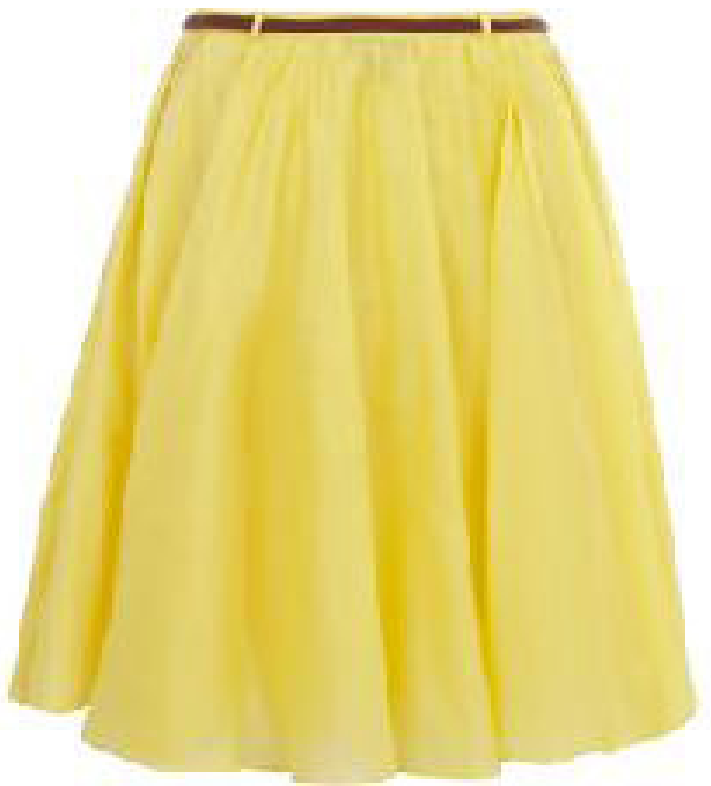} & \includegraphics[width=0.1\textwidth]{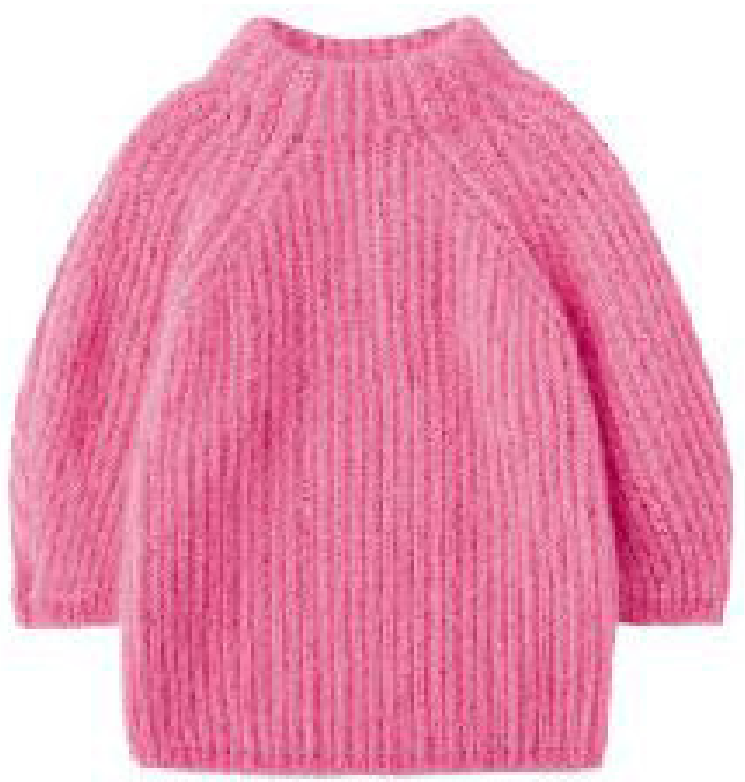} & \includegraphics[width=0.1\textwidth]{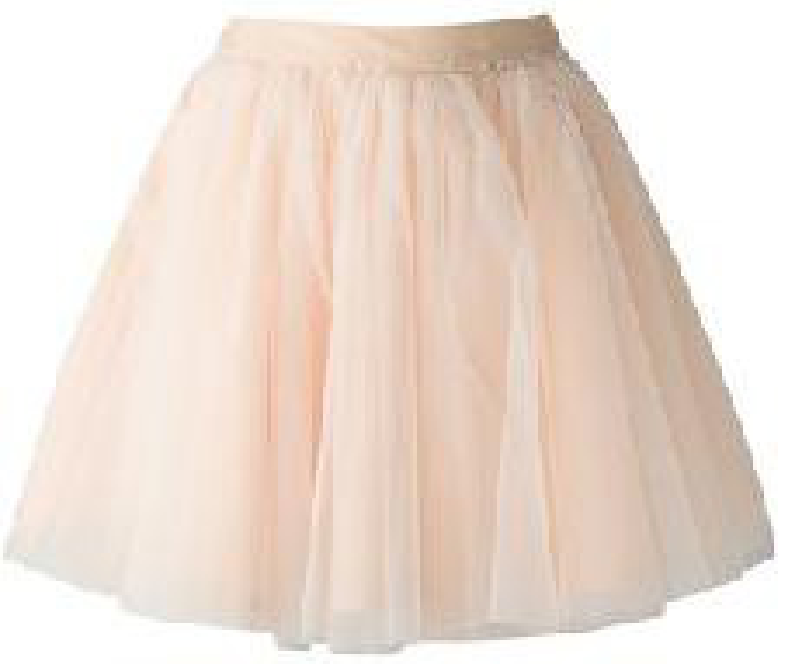} & \includegraphics[width=0.1\textwidth]{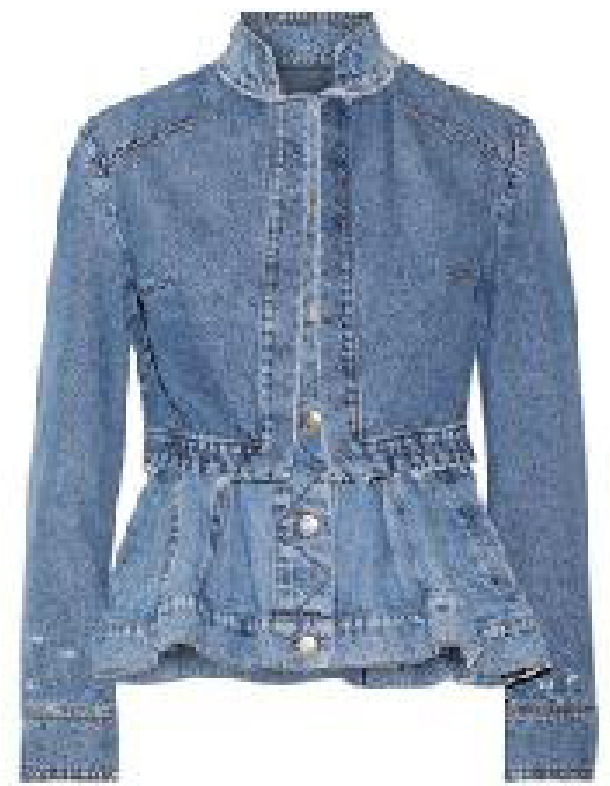} & \includegraphics[width=0.1\textwidth]{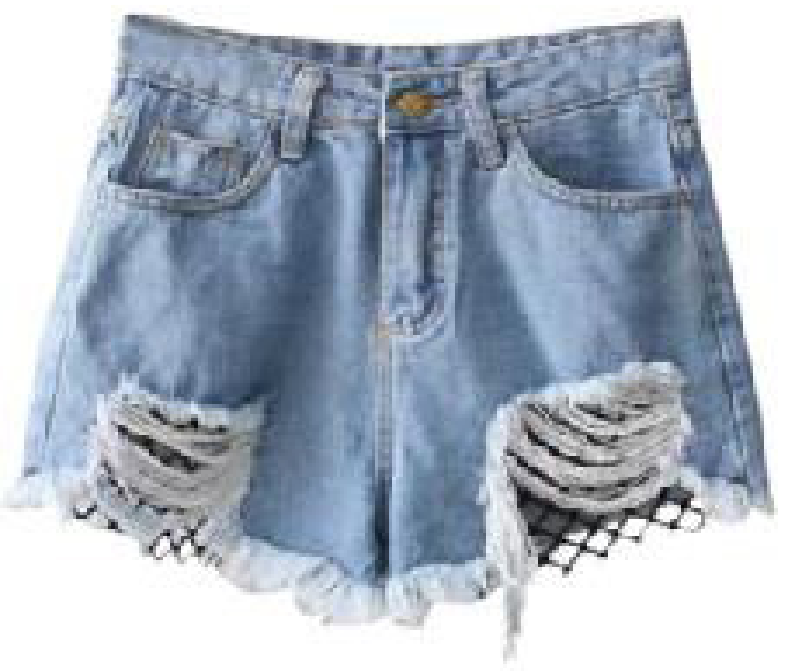} & \includegraphics[width=0.1\textwidth]{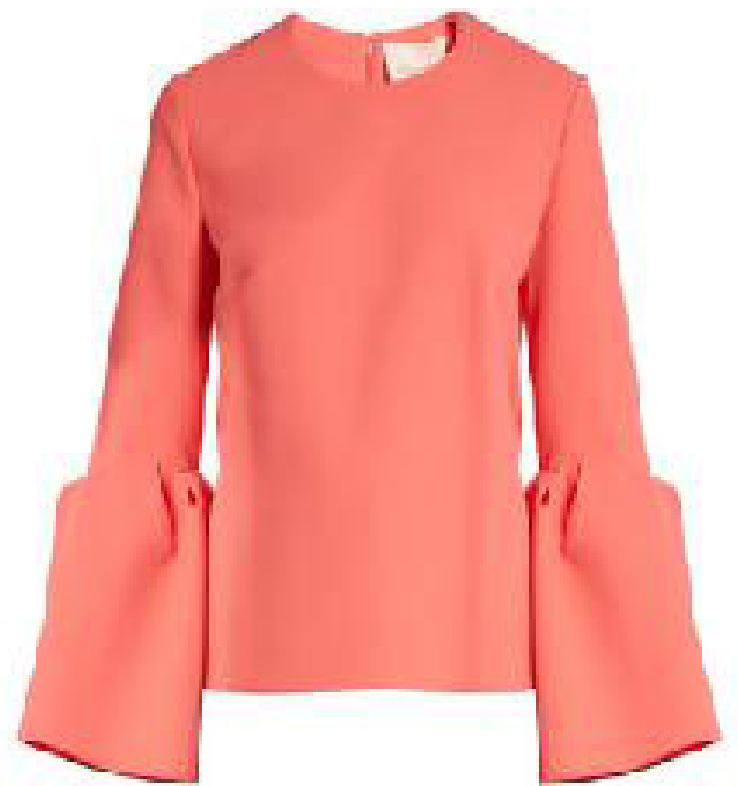} & \includegraphics[width=0.1\textwidth]{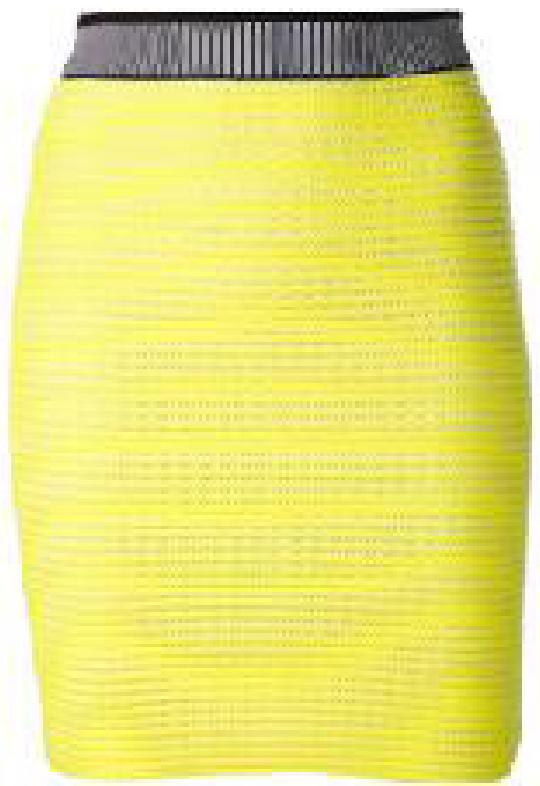} \\\hline\\[-1em]
\multicolumn{2}{p{4cm}|}{wow ! this is so beautiful ! love the skirt ! (\cmark)} & \multicolumn{2}{p{4cm}|}{love the pink ! (\cmark)} & \multicolumn{2}{p{4cm}|}{great denim look . (\cmark)} & \multicolumn{2}{p{4cm}}{love the color combination ! (\cmark)} \\\hline\\[-1em]
\includegraphics[width=0.1\textwidth]{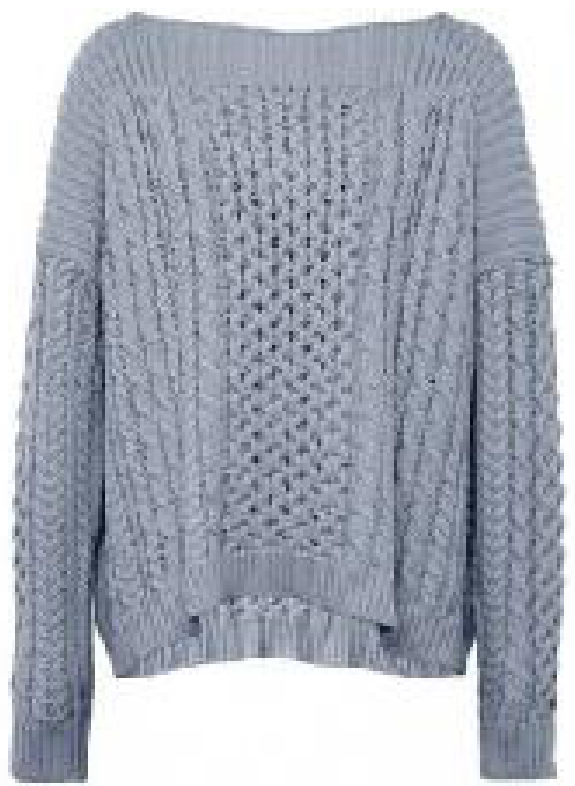} & \includegraphics[width=0.1\textwidth]{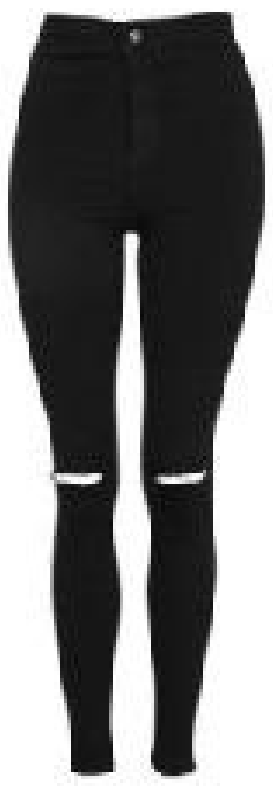} & \includegraphics[width=0.1\textwidth]{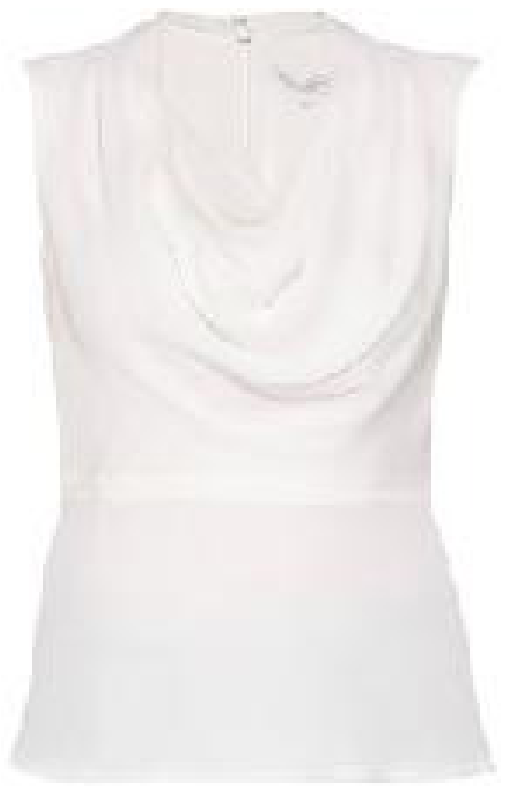} & \includegraphics[width=0.1\textwidth]{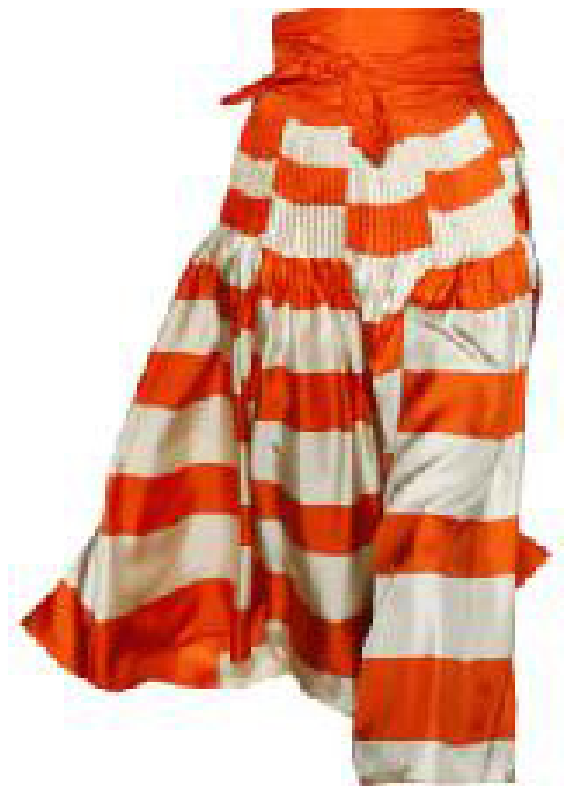} & \includegraphics[width=0.1\textwidth]{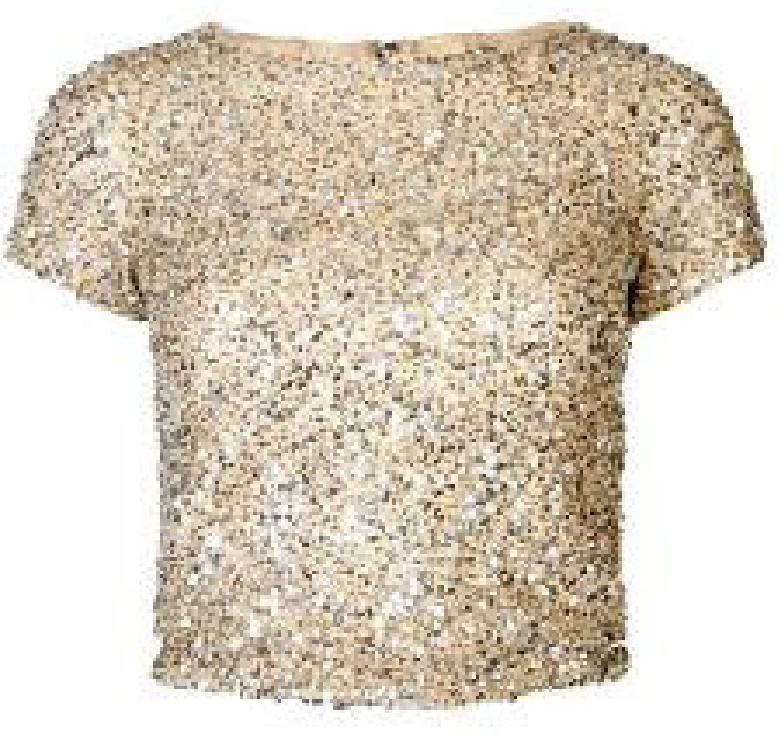} & \includegraphics[width=0.1\textwidth]{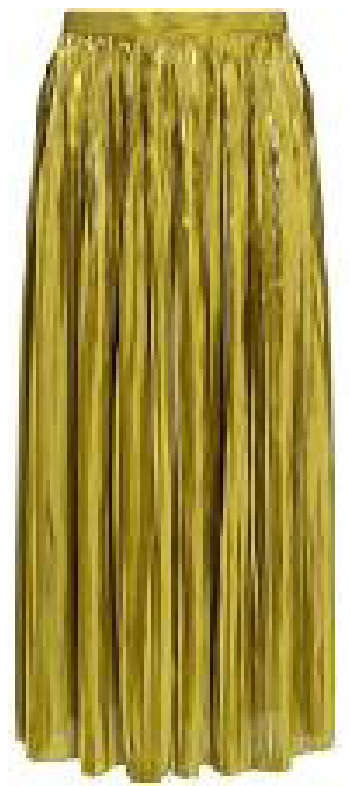} & \includegraphics[width=0.1\textwidth]{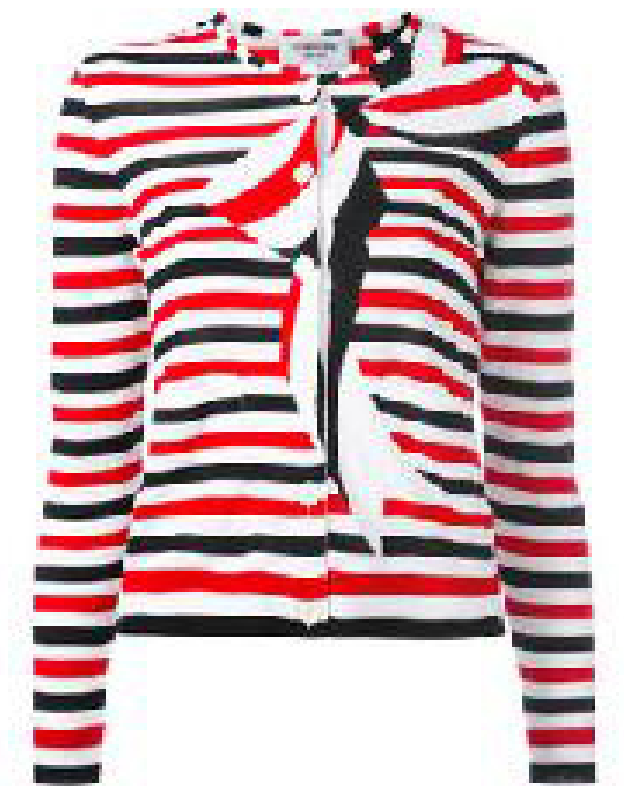} & \includegraphics[width=0.1\textwidth]{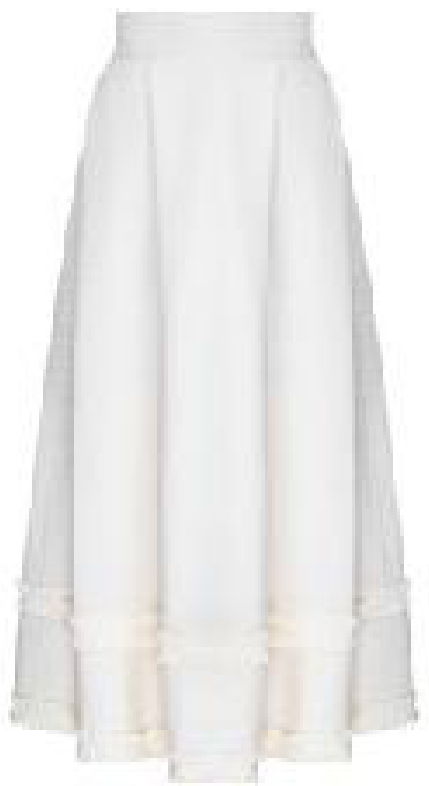} \\\hline\\[-1em]
\multicolumn{2}{p{4cm}|}{love this set ! the colours are amazing . (\cmark)} & \multicolumn{2}{p{4cm}|}{so beautiful and such a nice style here like it ~ . (\cmark)} & \multicolumn{2}{p{4cm}|}{great look great set great color . (\cmark)} & \multicolumn{2}{p{4cm}}{love the red and white ! (\cmark)} \\\hline\\[-1em]
\includegraphics[width=0.1\textwidth]{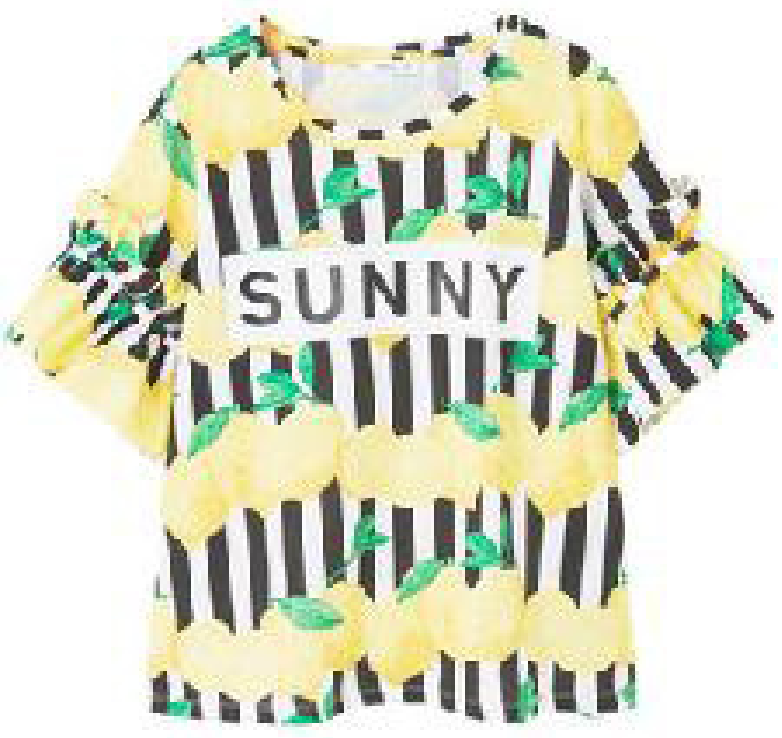} & \includegraphics[width=0.1\textwidth]{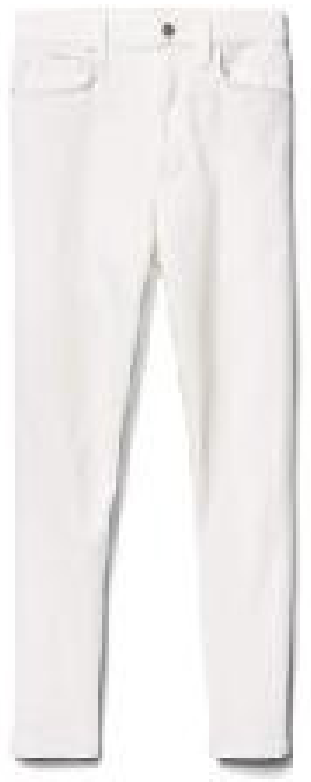} & \includegraphics[width=0.1\textwidth]{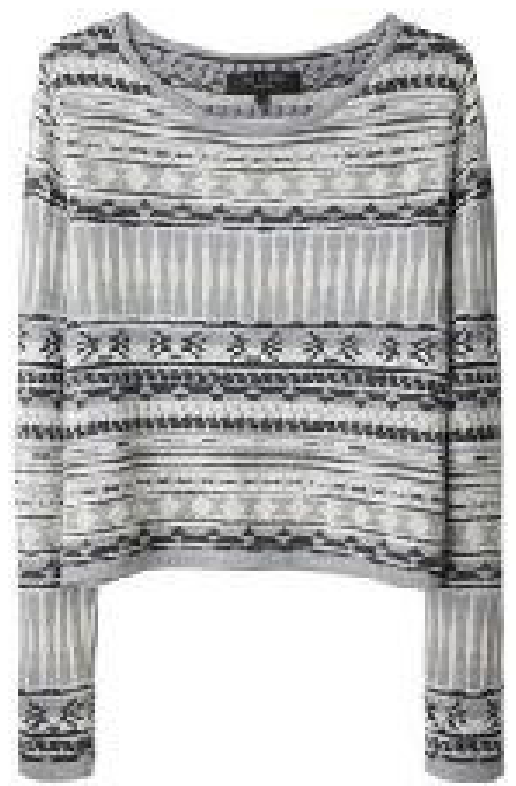} & \includegraphics[width=0.1\textwidth]{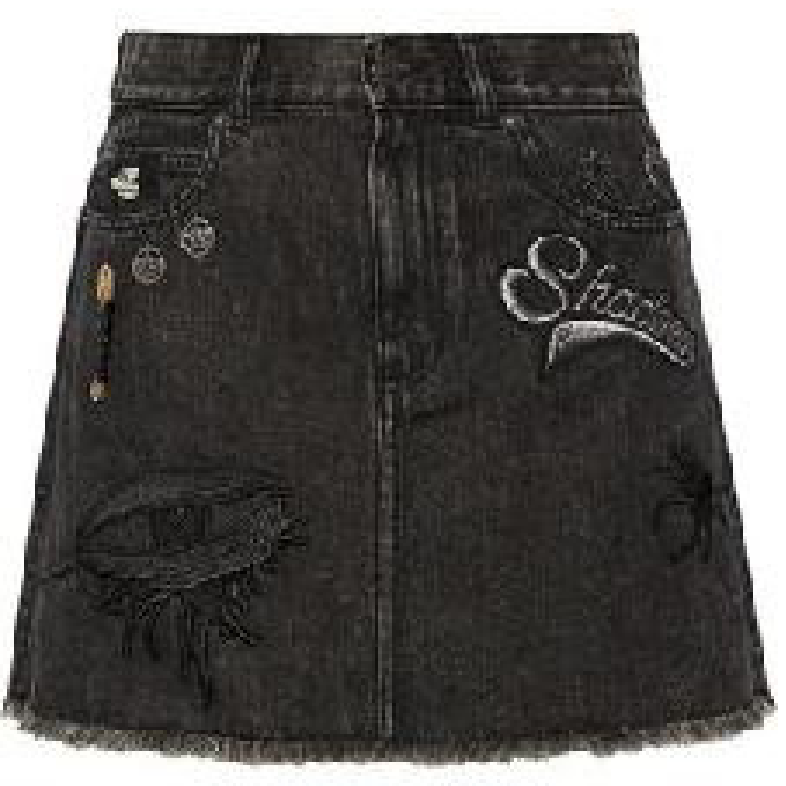} & \includegraphics[width=0.1\textwidth]{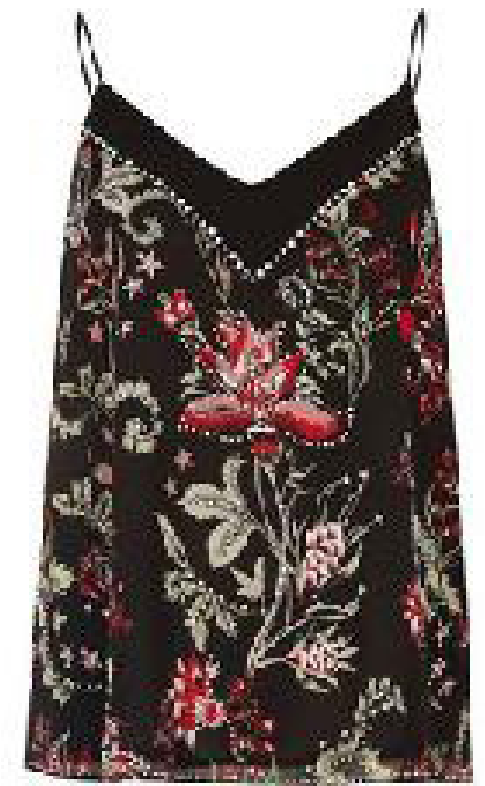} & \includegraphics[width=0.1\textwidth]{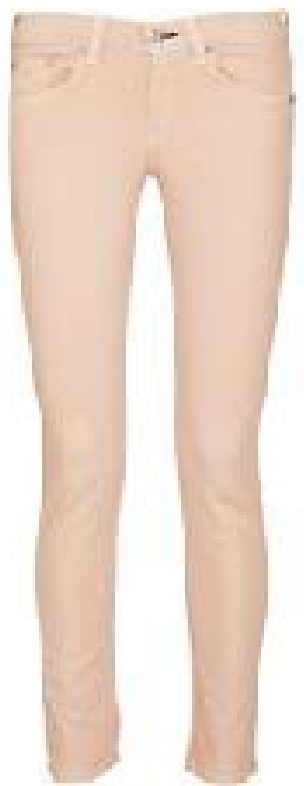} & \includegraphics[width=0.1\textwidth]{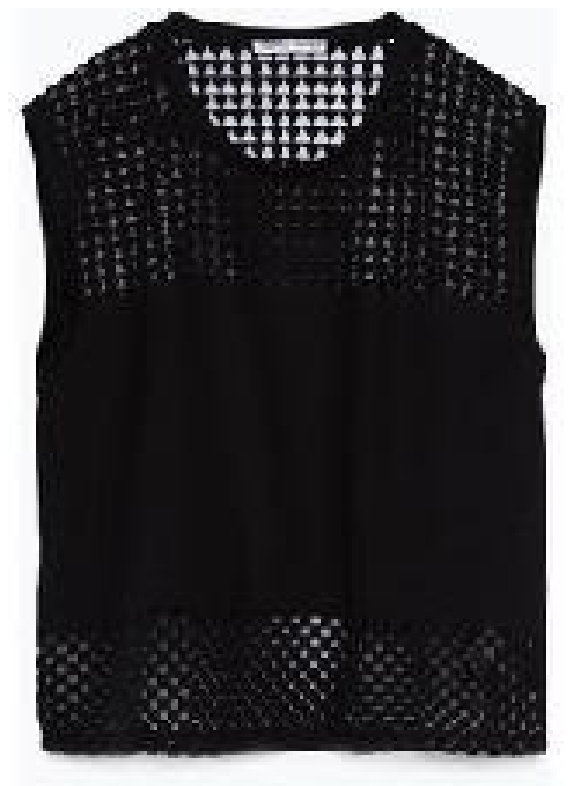} & \includegraphics[width=0.1\textwidth]{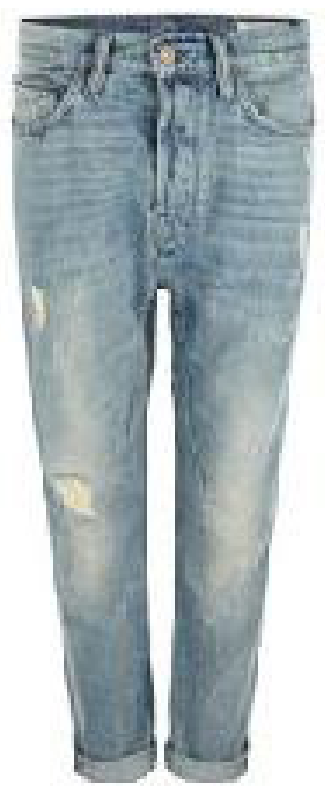} \\\hline\\[-1em]
\multicolumn{2}{p{4cm}|}{great look great set great mixing outfits n ' nice bag . (\xmark)} & \multicolumn{2}{p{4cm}|}{thank you so much for your lovely comments ! (\xmark)} & \multicolumn{2}{p{4cm}|}{congrats on top sets sweetie ! xxo . (\xmark)} & \multicolumn{2}{p{4cm}}{great set , love the shoes ! (\xmark)} \\\hline
\end{tabular}
\end{table*}

For the purpose of analyzing the linguistic quality of generated comments and the correlation between images and comments, we sample some instances from the test set, shown in Table \ref{ana_table5}.
We find that the generated comments are basically grammatical and syntactic.
And most of them express feelings and opinions about the combinations from the perspective of the public, which can be treated as explanations about why the top and the bottom match.
For example, ``wow! this is so beautiful! love the skirt!'' shows appreciation to this combination, and ``love the skirt" expresses a special preference for the skirt, which is also an appreciation of the outfit.
``Love the color combination" points out directly that color matching is the reason of the recommendation.
And ``so beautiful and such a nice style here i like it" expresses that the style of the outfit is beautiful and nice, which is a good explanation about why recommending this combination.
Additionally, \ac{NOR} generates comments like ``great denim look,'' where denim is the material of jeans and jackets.
Another example is ``love the pink," obviously because the top and the bottom are pink.
Similarly, ``love the red and white" finds that the top's color is red and the bottom's color is white.
In summary, \ac{NOR} is able to generate comments with visual features like texture, color and so on.

There are also some bad cases. 
For example, ``thank you so much for your lovely comments !", which is feedback on other users' comments, not a comment posted for the combination. In our datasets, a few comments are communications between users. 
This indicates that we should study better filtering methods in future work. 
Other bad cases include statements like ``nice bag''. In Polyvore, comments are for outfits, which include not only tops and bottoms, but also shoes, necklaces and so on. 
So generated comments may include items other than tops and bottoms. 
These bad cases imply that \ac{NOR} can generate words not only by visual features but also by ID or other information, which is confirmed when visualizing the effects of attention mechanisms in Section~\ref{section:giraffe}. 
There are some other problems we omit here, like duplicate comments or duplicate words, short comments and meaningless comments, which also push us to make further improvements.


\section{Conclusions and Future Work}

We have studied the task of explainable outfit recommendation. 
We have identified two main problems: the compatibility of fashion factors and the transformation between visual and textual information. 
To tackle these problems, we have proposed a deep learning-based framework, called \ac{NOR}, which simultaneously gives outfit recommendations and generates abstractive comments as explanations.
We have released a large real-world dataset, ExpFashion, including images, contextual metadata of items, and user comments.

In our experiments, we have demonstrated the effectiveness of \ac{NOR} and have found significant improvements over state-of-the-art baselines in terms of MAP, MRR and AUC.
Moreover, we have found that \ac{NOR} achieves impressive ROUGE and BLEU scores with respect to human-written comments.
We have also shown that the mutual attention and cross-modality attention mechanisms are useful for outfit recommendation and comment generation.

Limitations of our work include the fact that \ac{NOR} rarely generates negative comments to explain why an outfit does not match, that is because most of the comments in the dataset are positive.
Furthermore, as short comments take up a large percentage of the dataset, \ac{NOR} tends to generate short comments.

As to future work, we plan to explore more fashion items in our dataset, e.g., hats, glasses and shoes, etc.
Also, to alleviate the problem of generating meaningless comments, studies into coherence in information retrieval~\citep{he-effective-2009} or dialogue systems can be explored~\citep{N16-1014,vakulenko-measuring-2018}.
And we would like to incorporate other mechanisms, such as an auto-encoder, to further improve the performance.
Finally, we would like to build a personalized outfit recommendation system.

\ifCLASSOPTIONcompsoc
  \section*{Acknowledgments}
\else
  \section*{Acknowledgment}
\fi
We thank the anonymous reviewers for their helpful comments.

This work is supported by 
the Natural Science Foundation of China (61672324, 61672322), 
the Natural Science Foundation of Shandong province (2016ZRE27468), 
the Fundamental Research Funds of Shandong University,
Ahold Delhaize,
the Association of Universities in the Netherlands,
and
the Innovation Center for Artificial Intelligence (ICAI).
All content represents the opinion of the authors, which is not necessarily shared or endorsed by their respective employers and/or sponsors.

\ifCLASSOPTIONcaptionsoff
  \newpage
\fi

\bibliographystyle{IEEEtranN_changed}
\bibliography{tkde2018-fp-yujie-pengjie}

\begin{thebibliography}{69}
\providecommand{\natexlab}[1]{#1}
\providecommand{\url}[1]{#1}
\csname url@samestyle\endcsname
\providecommand{\newblock}{\relax}
\providecommand{\bibinfo}[2]{#2}
\providecommand{\BIBentrySTDinterwordspacing}{\spaceskip=0pt\relax}
\providecommand{\BIBentryALTinterwordstretchfactor}{4}
\providecommand{\BIBentryALTinterwordspacing}{\spaceskip=\fontdimen2\font plus
\BIBentryALTinterwordstretchfactor\fontdimen3\font minus
  \fontdimen4\font\relax}
\providecommand{\BIBforeignlanguage}[2]{{%
\expandafter\ifx\csname l@#1\endcsname\relax
\typeout{** WARNING: IEEEtranN.bst: No hyphenation pattern has been}%
\typeout{** loaded for the language `#1'. Using the pattern for}%
\typeout{** the default language instead.}%
\else
\language=\csname l@#1\endcsname
\fi
#2}}
\providecommand{\BIBdecl}{\relax}
\BIBdecl

\bibitem[Iwata et~al.(2011)Iwata, Watanabe, and Sawada]{Iwata2011}
Tomoharu Iwata, Shinji Watanabe, and Hiroshi Sawada, ``Fashion coordinates
  recommender system using photographs from fashion magazines,'' in
  \emph{International Joint Conference on Artificial Intelligence}, 2011, pp.
  2262--2267.

\bibitem[Liu et~al.(2012)Liu, Feng, Song, Zhang, Lu, Xu, and Yan]{Liu2012}
Si~Liu, Jiashi Feng, Zheng Song, Tianzhu Zhang, Hanqing Lu, Changsheng Xu, and
  Shuicheng Yan, ``Hi, magic closet, tell me what to wear!'' in \emph{ACM
  Multimedia}, 2012, pp. 619--628.

\bibitem[Tintarev and Masthoff(2007)]{tintarev2007survey}
Nava Tintarev and Judith Masthoff, ``A survey of explanations in recommender
  systems,'' in \emph{Data Engineering Workshop, 2007 IEEE 23rd International
  Conference on}.\hskip 1em plus 0.5em minus 0.4em\relax IEEE, 2007, pp.
  801--810.

\bibitem[Song et~al.(2017)Song, Feng, Liu, Li, Nie, and Ma]{Song2017}
Xuemeng Song, Fuli Feng, Jinhuan Liu, Zekun Li, Liqiang Nie, and Jun Ma,
  ``Neurostylist: Neural compatibility modeling for clothing matching,'' in
  \emph{ACM Multimedia}, 2017, pp. 753--761.

\bibitem[Hu et~al.(2015)Hu, Yi, and Davis]{Hu2015}
Yang Hu, Xi~Yi, and Larry~S. Davis, ``Collaborative fashion recommendation: A
  functional tensor factorization approach,'' in \emph{ACM Multimedia}, 2015,
  pp. 129--138.

\bibitem[Rendle et~al.(2009)Rendle, Freudenthaler, Gantner, and
  Schmidt-Thieme]{Rendle2009}
Steffen Rendle, Christoph Freudenthaler, Zeno Gantner, and Lars Schmidt-Thieme,
  ``Bpr: Bayesian personalized ranking from implicit feedback,'' in
  \emph{Conference on Uncertainty in Artificial Intelligence}, 2009, pp.
  452--461.

\bibitem[Jagadeesh et~al.(2014)Jagadeesh, Piramuthu, Bhardwaj, Di, and
  Sundaresan]{Jagadeesh2014}
Vignesh Jagadeesh, Robinson Piramuthu, Anurag Bhardwaj, Wei Di, and Neel
  Sundaresan, ``Large scale visual recommendations from street fashion
  images,'' in \emph{ACM SIGKDD Conference on Knowledge Discovery and Data
  Mining}, 2014, pp. 1925--1934.

\bibitem[McAuley et~al.(2015)McAuley, Targett, Shi, and van~den
  Hengel]{McAuley2015}
Julian McAuley, Christopher Targett, Qinfeng Shi, and Anton van~den Hengel,
  ``Image-based recommendations on styles and substitutes,'' in
  \emph{International Conference on Research and Development in Information
  Retrieval}, 2015, pp. 43--52.

\bibitem[He and McAuley(2016)]{He2016}
Ruining He and Julian McAuley, ``{VBPR}: Visual bayesian personalized ranking
  from implicit feedback,'' in \emph{AAAI Conference on Artificial
  Intelligence}, 2016, pp. 144--150.

\bibitem[Li et~al.(2017{\natexlab{a}})Li, Cao, Zhu, and Luo]{Li2017MiningFO}
Yuncheng Li, Liangliang Cao, Jiang Zhu, and Jiebo Luo, ``Mining fashion outfit
  composition using an end-to-end deep learning approach on set data,'' in
  \emph{IEEE Transactions on Multimedia}, vol.~19.\hskip 1em plus 0.5em minus
  0.4em\relax IEEE, 2017, pp. 1946--1955.

\bibitem[Kang et~al.(2017)Kang, Fang, Wang, and McAuley]{Kang2017Visually}
Wang~Cheng Kang, Chen Fang, Zhaowen Wang, and Julian McAuley, ``Visually-aware
  fashion recommendation and design with generative image models,'' in
  \emph{International Conference on Data Mining}, 2017, pp. 207--216.

\bibitem[Han et~al.(2017)Han, Wu, Jiang, and Davis]{Han2017}
Xintong Han, Zuxuan Wu, Yu-Gang Jiang, and Larry~S. Davis, ``Learning fashion
  compatibility with bidirectional lstms,'' in \emph{ACM Multimedia}, 2017, pp.
  1078--1086.

\bibitem[Song et~al.(2018)Song, Feng, Han, Yang, Liu, and Nie]{Song2018Neural}
Xuemeng Song, Fuli Feng, Xianjing Han, Xin Yang, Wei Liu, and Liqiang Nie,
  ``Neural compatibility modeling with attentive knowledge distillation,'' in
  \emph{International Conference on Research on Development in Information
  Retrieval (SIGIR'18)}, 2018.

\bibitem[Vig et~al.(2009)Vig, Sen, and Riedl]{Vig2009}
Jesse Vig, Shilad Sen, and John Riedl, ``Tagsplanations: Explaining
  recommendations using tags,'' in \emph{International Conference on
  Intelligent User Interfaces}, 2009, pp. 47--56.

\bibitem[Zhang et~al.(2014)Zhang, Lai, Zhang, Zhang, Liu, and Ma]{Zhang2014}
Yongfeng Zhang, Guokun Lai, Min Zhang, Yi~Zhang, Yiqun Liu, and Shaoping Ma,
  ``Explicit factor models for explainable recommendation based on phrase-level
  sentiment analysis,'' in \emph{International Conference on Research and
  Development in Information Retrieval}, 2014, pp. 83--92.

\bibitem[He et~al.(2015)He, Chen, Kan, and Chen]{He2015}
Xiangnan He, Tao Chen, Min-Yen Kan, and Xiao Chen, ``Trirank: Review-aware
  explainable recommendation by modeling aspects,'' in \emph{ACM International
  Conference on Information and Knowledge Management}, 2015, pp. 1661--1670.

\bibitem[Ribeiro et~al.(2016)Ribeiro, Singh, and Guestrin]{Ribeiro2016}
Marco~Tulio Ribeiro, Sameer Singh, and Carlos Guestrin, ````{Why} should {I}
  trust you?'': Explaining the predictions of any classifier,'' in \emph{ACM
  SIGKDD Conference on Knowledge Discovery and Data Mining}, 2016, pp.
  1135--1144.

\bibitem[Ren et~al.(2017)Ren, Liang, Li, Wang, and de~Rijke]{Ren2017}
Zhaochun Ren, Shangsong Liang, Piji Li, Shuaiqiang Wang, and Maarten de~Rijke,
  ``Social collaborative viewpoint regression with explainable
  recommendations,'' in \emph{International Conference on Web Search and Data
  Mining}, 2017, pp. 485--494.

\bibitem[Wang et~al.(2018)Wang, Wang, Jia, and Yin]{wang2018explainable}
Nan Wang, Hongning Wang, Yiling Jia, and Yue Yin, ``Explainable recommendation
  via multi-task learning in opinionated text data,'' in \emph{International
  Conference on Research and Development in Information Retrieval}, 2018.

\bibitem[Ni et~al.(2017)Ni, Lipton, Vikram, and McAuley]{NiLipVikMcA17}
Jianmo Ni, Zachary Lipton, Sharad Vikram, and Julian McAuley, ``Estimating
  reactions and recommending products with generative models of reviews,'' in
  \emph{International Joint Conference on Natural Language Processing}, 2017,
  pp. 783--791.

\bibitem[Li et~al.(2017{\natexlab{b}})Li, Wang, Ren, Bing, and Lam]{Li2017}
Piji Li, Zihao Wang, Zhaochun Ren, Lidong Bing, and Wai Lam, ``Neural rating
  regression with abstractive tips generation for recommendation,'' in
  \emph{International Conference on Research and Development in Information
  Retrieval}, 2017, pp. 345--354.

\bibitem[Li et~al.(2017{\natexlab{c}})Li, Lam, Bing, and Wang]{Li2017Deep}
Piji Li, Wai Lam, Lidong Bing, and Zihao Wang, ``Deep recurrent generative
  decoder for abstractive text summarization,'' in \emph{Conference on
  Empirical Methods in Natural Language Processing}, 2017, pp. 2091--2100.

\bibitem[Zhou et~al.(2017)Zhou, Yang, Wei, and Zhou]{Zhou2017Selective}
Qingyu Zhou, Nan Yang, Furu Wei, and Ming Zhou, ``Selective encoding for
  abstractive sentence summarization,'' in \emph{Annual Meeting of the
  Association for Computational Linguistics}, 2017, pp. 1095--1104.

\bibitem[Bahdanau et~al.(2015)Bahdanau, Cho, and Bengio]{Bahdanau2014Neural}
Dzmitry Bahdanau, Kyunghyun Cho, and Yoshua Bengio, ``Neural machine
  translation by jointly learning to align and translate,'' in
  \emph{International Conference on Learning Representations}, 2015.

\bibitem[Vaswani et~al.(2017)Vaswani, Shazeer, Parmar, Uszkoreit, Jones, Gomez,
  Kaiser, and Polosukhin]{vaswani2017attention}
Ashish Vaswani, Noam Shazeer, Niki Parmar, Jakob Uszkoreit, Llion Jones,
  Aidan~N Gomez, Lukasz Kaiser, and Illia Polosukhin, ``Attention is all you
  need,'' in \emph{Advances in Neural Information Processing Systems}, 2017,
  pp. 5998--6008.

\bibitem[Serban et~al.(2016)Serban, Sordoni, Bengio, Courville, and
  Pineau]{Serban2016Building}
Iulian~V. Serban, Alessandro Sordoni, Yoshua Bengio, Aaron Courville, and
  Joelle Pineau, ``Building end-to-end dialogue systems using generative
  hierarchical neural network models,'' in \emph{AAAI Conference on Artificial
  Intelligence}, 2016, pp. 3776--3783.

\bibitem[Williams et~al.(2017)Williams, Asadi, and Zweig]{Williams2017Hybrid}
Jason~D. Williams, Kavosh Asadi, and Geoffrey Zweig, ``Hybrid code networks:
  practical and efficient end-to-end dialog control with supervised and
  reinforcement learning,'' in \emph{Annual Meeting of the Association for
  Computational Linguistics}, 2017, pp. 665--677.

\bibitem[Xu et~al.(2015)Xu, Ba, Kiros, Cho, Courville, Salakhudinov, Zemel, and
  Bengio]{Xu2015Show}
Kelvin Xu, Jimmy Ba, Ryan Kiros, Kyunghyun Cho, Aaron Courville, Ruslan
  Salakhudinov, Rich Zemel, and Yoshua Bengio, ``Show, attend and tell: Neural
  image caption generation with visual attention,'' in \emph{International
  Conference on Machine Learning}, 2015, pp. 2048--2057.

\bibitem[Chen et~al.(2017)Chen, Zhang, Xiao, Nie, Shao, Liu, and
  Chua]{Chen2017SCA}
Long Chen, Hanwang Zhang, Jun Xiao, Liqiang Nie, Jian Shao, Wei Liu, and
  Tat~Seng Chua, ``Sca-cnn: Spatial and channel-wise attention in convolutional
  networks for image captioning,'' in \emph{IEEE Conference on Computer Vision
  and Pattern Recognition}, 2017, pp. 6298--6306.

\bibitem[Cao et~al.(2012)Cao, Chen, Long, Zheng, and Yu]{Cao2012News}
Xuezhi Cao, Kailong Chen, Rui Long, Guoqing Zheng, and Yong Yu, ``News comments
  generation via mining microblogs,'' in \emph{International World Wide Web
  Conference}, 2012, pp. 471--472.

\bibitem[Lipton et~al.(2015)Lipton, Vikram, and McAuley]{lipton2015capturing}
\BIBentryALTinterwordspacing
Zachary~C Lipton, Sharad Vikram, and Julian McAuley, ``Capturing meaning in
  product reviews with character-level generative text models,'' \emph{CoRR},
  2015. [Online]. Available: \url{http://arxiv.org/abs/1511.03683}
\BIBentrySTDinterwordspacing

\bibitem[Radford et~al.(2017)Radford, Jozefowicz, and
  Sutskever]{radford2017learning}
\BIBentryALTinterwordspacing
Alec Radford, Rafal Jozefowicz, and Ilya Sutskever, ``Learning to generate
  reviews and discovering sentiment,'' \emph{CoRR}, 2017. [Online]. Available:
  \url{https://arxiv.org/abs/1704.01444}
\BIBentrySTDinterwordspacing

\bibitem[Dong et~al.(2017)Dong, Huang, Wei, Lapata, Zhou, and
  Xu]{Dong2017Learning}
Li~Dong, Shaohan Huang, Furu Wei, Mirella Lapata, Ming Zhou, and Ke~Xu,
  ``Learning to generate product reviews from attributes,'' in \emph{Conference
  of the European Chapter of the Association for Computational Linguistics},
  2017, pp. 623--632.

\bibitem[Tang et~al.(2016)Tang, Yang, Carton, Zhang, and Mei]{Tang2016Context}
Jian Tang, Yifan Yang, Sam Carton, Ming Zhang, and Qiaozhu Mei, ``Context-aware
  natural language generation with recurrent neural networks,'' in \emph{arXiv
  preprint arXiv:1611.09900}, 2016.

\bibitem[Hu et~al.(2017)Hu, Yang, Liang, Salakhutdinov, and
  Xing]{Hu2017Controllable}
Zhiting Hu, Zichao Yang, Xiaodan Liang, Ruslan Salakhutdinov, and Eric~P. Xing,
  ``Controllable text generation,'' in \emph{International Conference on
  Machine Learning}, 2017, pp. 1587--1596.

\bibitem[LeCun et~al.(1998)LeCun, Bottou, Bengio, and
  Haffner]{Lecun1998Gradient}
Yann LeCun, L{\'e}on Bottou, Yoshua Bengio, and Patrick Haffner,
  ``Gradient-based learning applied to document recognition,'' in
  \emph{Proceedings of the IEEE}, vol.~86, no.~11.\hskip 1em plus 0.5em minus
  0.4em\relax IEEE, 1998, pp. 2278--2324.

\bibitem[Cho et~al.(2014)Cho, Merrienboer, Bahdanau, and Bengio]{Cho2014On}
Kyunghyun Cho, Bart~Van Merrienboer, Dzmitry Bahdanau, and Yoshua Bengio, ``On
  the properties of neural machine translation: Encoder-decoder approaches,''
  in \emph{Proceedings of the 8th Workshop on Syntax, Semantics and Structure
  in Statistical Translation}, 2014, pp. 103--111.

\bibitem[He et~al.(2016)He, Zhang, Ren, and Sun]{He2016Deep}
Kaiming He, Xiangyu Zhang, Shaoqing Ren, and Jian Sun, ``Deep residual learning
  for image recognition,'' in \emph{IEEE Conference on Computer Vision and
  Pattern Recognition}, 2016, pp. 770--778.

\bibitem[Huang et~al.(2017)Huang, Liu, van~der Maaten, and
  Weinberger]{huang2017densely}
Gao Huang, Zhuang Liu, Laurens van~der Maaten, and Kilian~Q Weinberger,
  ``Densely connected convolutional networks,'' in \emph{IEEE Conference on
  Computer Vision and Pattern Recognition}, 2017, pp. 4700--4708.

\bibitem[Luong et~al.(2015)Luong, Pham, and Manning]{Luong2015Effective}
Minh~Thang Luong, Hieu Pham, and Christopher~D Manning, ``Effective approaches
  to attention-based neural machine translation,'' in \emph{Empirical Methods
  on Natural Language Processing}, 2015, pp. 1412--1421.

\bibitem[Koren et~al.(2009)Koren, Bell, and Volinsky]{Koren2009Matrix}
Yehuda Koren, Robert Bell, and Chris Volinsky, ``Matrix factorization
  techniques for recommender systems,'' in \emph{IEEE Computer Society Press},
  vol.~42, no.~8.\hskip 1em plus 0.5em minus 0.4em\relax IEEE, 2009, pp.
  30--37.

\bibitem[Lee and Seung(2000)]{Lee2000Algorithms}
Daniel~D. Lee and H.~Sebastian Seung, ``Algorithms for non-negative matrix
  factorization,'' in \emph{Annual Conference on Neural Information Processing
  Systems}, 2000, pp. 535--541.

\bibitem[Salakhutdinov and Mnih(2007)]{Salakhutdinov2007Probabilistic}
Ruslan Salakhutdinov and Andriy Mnih, ``Probabilistic matrix factorization,''
  in \emph{Annual Conference on Neural Information Processing Systems}, 2007,
  pp. 1257--1264.

\bibitem[Kiapour et~al.(2015)Kiapour, Han, Lazebnik, Berg, and Berg]{7410739}
M.~Hadi Kiapour, Xufeng Han, Svetlana Lazebnik, Alexander~C. Berg, and
  Tamara~L. Berg, ``Where to buy it: Matching street clothing photos in online
  shops,'' in \emph{IEEE International Conference on Computer Vision}, 2015,
  pp. 3343--3351.

\bibitem[Yamaguchi et~al.(2012)Yamaguchi, Kiapour, E.Ortiz, and
  L.~Berg]{Yamaguchi2012}
Kota Yamaguchi, M.~Hadi Kiapour, Luis E.Ortiz, and Tamara L.~Berg, ``Parsing
  clothing in fashion photographs,'' in \emph{IEEE Conference on Computer
  Vision and Pattern Recognition}, 2012, pp. 3570--3577.

\bibitem[Yamaguchi et~al.(2015)Yamaguchi, Kiapour, Ortiz, and Berg]{6888484}
Kota Yamaguchi, M.~Hadi Kiapour, Luis~E. Ortiz, and Tamara~L. Berg,
  ``Retrieving similar styles to parse clothing,'' in \emph{IEEE Transactions
  on Pattern Analysis and Machine Intelligence}, vol.~37, no.~5.\hskip 1em plus
  0.5em minus 0.4em\relax IEEE, 2015, pp. 1028--1040.

\bibitem[Glorot and Bengio(2010)]{Glorot2010Understanding}
Xavier Glorot and Yoshua Bengio, ``Understanding the difficulty of training
  deep feedforward neural networks,'' in \emph{Journal of Machine Learning
  Research}, vol.~9, 2010, pp. 249--256.

\bibitem[Kingma and Ba(2015)]{Kingma2014Adam}
\BIBentryALTinterwordspacing
Diederik~P. Kingma and Jimmy Ba, ``Adam: A method for stochastic
  optimization,'' in \emph{International Conference on Learning
  Representations}, 2015. [Online]. Available:
  \url{http://arxiv.org/abs/1412.6980}
\BIBentrySTDinterwordspacing

\bibitem[Pascanu et~al.(2013)Pascanu, Mikolov, and Bengio]{Pascanu2013}
Razvan Pascanu, Tomas Mikolov, and Yoshua Bengio, ``On the difficulty of
  training recurrent neural networks,'' in \emph{International Conference on
  Machine Learning}, 2013, pp. III--1310--III--1318.

\bibitem[Koehn(2004)]{Koehn2004Pharaoh}
Philipp Koehn, ``Pharaoh: A beam search decoder for phrase-based statistical
  machine translation models,'' in \emph{Association for Machine Translation in
  the Americas}, 2004, pp. 115--124.

\bibitem[Abadi et~al.(2015)Abadi, Agarwal, Barham, Brevdo, Chen, Citro,
  Corrado, Davis, Dean, Devin, et~al.]{Abadi2015TensorFlow}
Mart{\'\i}n Abadi, Ashish Agarwal, Paul Barham, Eugene Brevdo, Zhifeng Chen,
  Craig Citro, Greg Corrado, Andy Davis, Jeffrey Dean, Matthieu Devin
  \emph{et~al.}, ``Tensorflow: Large-scale machine learning on heterogeneous
  distributed systems,'' in \emph{CoRR}, vol. abs/1603.04467, 2015.

\bibitem[He et~al.(2017)He, Liao, Zhang, Nie, Hu, and Chua]{He2017NCF}
Xiangnan He, Lizi Liao, Hanwang Zhang, Liqiang Nie, Xia Hu, and Tat-Seng Chua,
  ``Neural collaborative filtering,'' in \emph{International World Wide Web
  Conference}, 2017, pp. 173--182.

\bibitem[Chatfield et~al.(2014)Chatfield, Simonyan, Vedaldi, and
  Zisserman]{Chatfield14}
Ken Chatfield, Karen Simonyan, Andrea Vedaldi, and Andrew Zisserman, ``Return
  of the devil in the details: Delving deep into convolutional nets,'' in
  \emph{British Machine Vision Conference}, 2014.

\bibitem[Jia et~al.(2014)Jia, Shelhamer, Donahue, Karayev, Long, Girshick,
  Guadarrama, and Darrell]{Jia2014}
Yangqing Jia, Evan Shelhamer, Jeff Donahue, Sergey Karayev, Jonathan Long, Ross
  Girshick, Sergio Guadarrama, and Trevor Darrell, ``Caffe: Convolutional
  architecture for fast feature embedding,'' in \emph{ACM Multimedia}.\hskip
  1em plus 0.5em minus 0.4em\relax ACM, 2014, pp. 675--678.

\bibitem[Aharon et~al.(2015)Aharon, Anava, Avigdor-Elgrabli, Drachsler-Cohen,
  Golan, and Somekh]{Aharon2015ExcUseMe}
Michal Aharon, Oren Anava, Noa Avigdor-Elgrabli, Dana Drachsler-Cohen, Shahar
  Golan, and Oren Somekh, ``Excuseme:asking users to help in item cold-start
  recommendations,'' in \emph{ACM Conference on Recommender Systems}, 2015, pp.
  83--90.

\bibitem[Rendle and Schmidt-Thieme(2010)]{Rendle2010}
Steffen Rendle and Lars Schmidt-Thieme, ``Pairwise interaction tensor
  factorization for personalized tag recommendation,'' in \emph{International
  Conference on Web Search and Data Mining}, 2010, pp. 81--90.

\bibitem[Zhang et~al.(2013)Zhang, Zha, Yang, Yan, Gao, and Chua]{Zhang2013}
Hanwang Zhang, Zheng-Jun Zha, Yang Yang, Shuicheng Yan, Yue Gao, and Tat-Seng
  Chua, ``Attribute-augmented semantic hierarchy: Towards bridging semantic gap
  and intention gap in image retrieval,'' in \emph{ACM Multimedia}, 2013, pp.
  33--42.

\bibitem[Li et~al.(2017{\natexlab{d}})Li, Ren, Chen, Ren, Lian, and
  Ma]{DBLP:conf/cikm/LiRCRLM17}
Jing Li, Pengjie Ren, Zhumin Chen, Zhaochun Ren, Tao Lian, and Jun Ma, ``Neural
  attentive session-based recommendation,'' in \emph{ACM International
  Conference on Information and Knowledge Management}, 2017, pp. 1419--1428.

\bibitem[Qian et~al.(2014)Qian, Feng, Zhao, and Mei]{Qian2014Personalized}
Xueming Qian, He~Feng, Guoshuai Zhao, and Tao Mei, ``Personalized
  recommendation combining user interest and social circle,'' in \emph{IEEE
  Transactions on Knowledge \& Data Engineering}, vol.~26, no.~7, 2014, pp.
  1763--1777.

\bibitem[Wang and Wang(2014)]{Wang2014Improving}
Xinxi Wang and Ye~Wang, ``Improving content-based and hybrid music
  recommendation using deep learning,'' in \emph{ACM Multimedia}, 2014, pp.
  627--636.

\bibitem[Erkan and Radev(2004)]{Erkan2004}
G\"{u}nes Erkan and Dragomir~R. Radev, ``Lexrank: Graph-based lexical
  centrality as salience in text summarization,'' in \emph{Journal of
  Artificial Intelligence Research}, vol.~22, no.~1.\hskip 1em plus 0.5em minus
  0.4em\relax AI Access Foundation, 2004, pp. 457--479.

\bibitem[Wang and Blei(2011)]{Wang2011}
Chong Wang and David~M. Blei, ``Collaborative topic modeling for recommending
  scientific articles,'' in \emph{ACM SIGKDD Conference on Knowledge Discovery
  and Data Mining}, 2011, pp. 448--456.

\bibitem[Ling et~al.(2014)Ling, Lyu, and King]{Ling2014}
Guang Ling, Michael~R. Lyu, and Irwin King, ``Ratings meet reviews, a combined
  approach to recommend,'' in \emph{ACM RecSys}, 2014, pp. 105--112.

\bibitem[Lin(2004)]{Lin2004}
Chin-Yew Lin, ``Rouge: a package for automatic evaluation of summaries,'' in
  \emph{Workshop on Text Summarization Branches Out, The Association for
  Computational Linguistics}, 2004.

\bibitem[Papineni et~al.(2002)Papineni, Roukos, Ward, and Zhu]{Papineni2002}
Kishore Papineni, Salim Roukos, Todd Ward, and Wei-Jing Zhu, ``Bleu: A method
  for automatic evaluation of machine translation,'' in \emph{The Association
  for Computational Linguistics}, 2002, pp. 311--318.

\bibitem[Caruana(1998)]{Caruana1997}
Rich Caruana, ``Multitask learning,'' in \emph{Learning to learn}.\hskip 1em
  plus 0.5em minus 0.4em\relax Springer, 1998, pp. 95--133.

\bibitem[He et~al.(2009)He, Weerkamp, Larson, and de~Rijke]{he-effective-2009}
Jiyin He, Wouter Weerkamp, Martha Larson, and Maarten de~Rijke, ``An effective
  coherence measure to determine topical consistency in user generated
  content,'' \emph{International Journal on Document Analysis and Recognition},
  vol.~12, no.~3, pp. 185--203, September 2009.

\bibitem[Li et~al.(2016)Li, Galley, Brockett, Gao, and Dolan]{N16-1014}
Jiwei Li, Michel Galley, Chris Brockett, Jianfeng Gao, and Bill Dolan, ``A
  diversity-promoting objective function for neural conversation models,'' in
  \emph{The North American Chapter of the Association for Computational
  Linguistics: Human Language Technologies}, 2016, pp. 110--119.

\bibitem[Vakulenko et~al.(2018)Vakulenko, de~Rijke, Cochez, Savenkov, and
  Polleres]{vakulenko-measuring-2018}
Svitlana Vakulenko, Maarten de~Rijke, Michael Cochez, Vadim Savenkov, and Axel
  Polleres, ``Measuring semantic coherence of a conversation,'' in \emph{ISWC
  2018: 17th International Semantic Web Conference}.\hskip 1em plus 0.5em minus
  0.4em\relax Springer, October 2018, pp. 634--651.

\end{thebibliography}

\newpage
\begin{IEEEbiography}[{\includegraphics[width=1in,height=1.25in,clip,keepaspectratio]{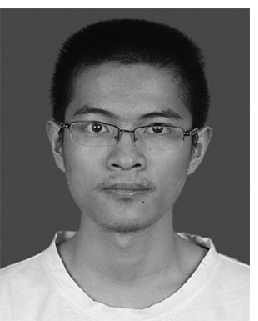}}]{Yujie Lin}
received B.S. from Shandong University, in 2016. Currently he is a master in Shandong University, supervised by Jun Ma. His research area is in information retrieval, recommender system and deep learning.
\end{IEEEbiography}
\vspace{-5em}
\begin{IEEEbiography}[{\includegraphics[width=1in,height=1.25in,clip,keepaspectratio]{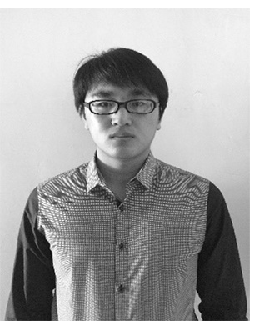}}]{Pengjie Ren}
is a postdoc researcher in the Informatics Institute, University of Amsterdam, Amsterdam, The Netherlands. His research interests fall in information retrieval, natural language processing, and recommender systems. He has previously published at TOIS, SIGIR, AAAI, CIKM, and COLING.
\end{IEEEbiography}
\vspace{-5em}
\begin{IEEEbiography}[{\includegraphics[width=1in,height=1.25in,clip,keepaspectratio]{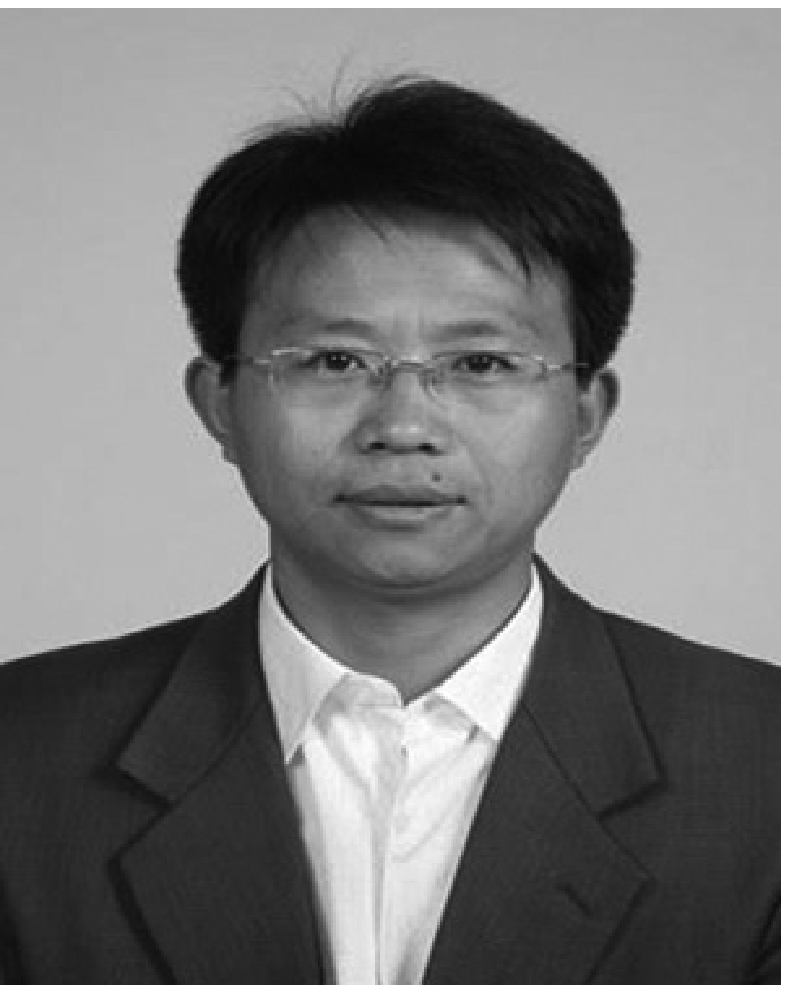}}]{Zhumin Chen}
is an associate professor in School of Computer Science and Technology of Shandong University. He is a member of the Chinese Information Technology Committee, Social Media Processing Committee, China Computer Federation Technical Committee (CCF) and ACM. He received his Ph.D. from Shandong University. His research interests mainly include information retrieval, big data mining and processing, as well as social media processing.
\end{IEEEbiography}
\vspace{-5em}
\begin{IEEEbiography}[{\includegraphics[width=1in,height=1.25in,clip,keepaspectratio]{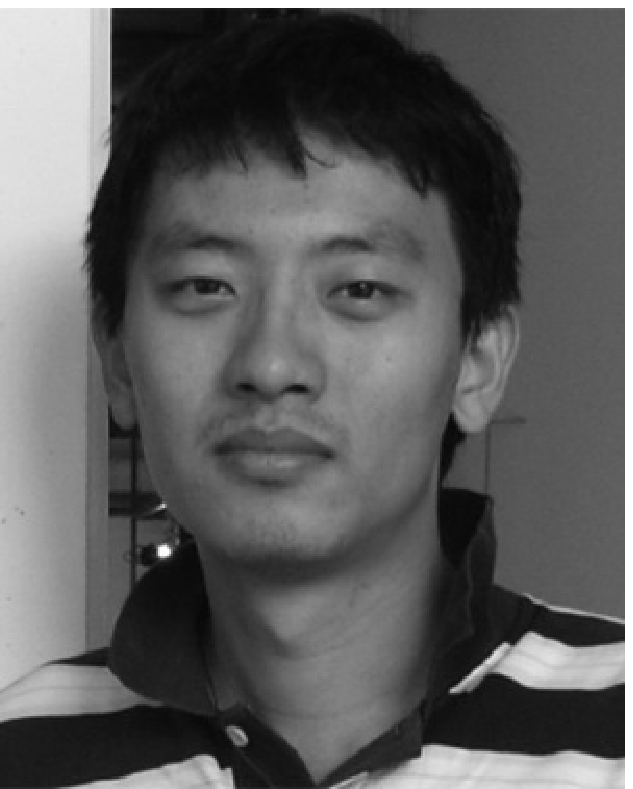}}]{Zhaochun Ren}
received his MSc degree from Shandong University in 2012, and a PhD degree from the University of Amsterdam in 2016. He is a professor at Shandong University. He previously was a research scientist in JD.com. Before that he worked as a research associate in University of London. He also worked as a short-term visiting scholar in Max-Planck-Institut f\"{u}r Informatik in 2012. He is interested in information retrieval, natural language processing, social media mining, and content analysis in e-discovery. He has previously published at SIGIR, ACL, WSDM, CIKM, and KDD.
\end{IEEEbiography}
\vspace{-5em}
\begin{IEEEbiography}[{\includegraphics[width=1in,height=1.25in,clip,keepaspectratio]{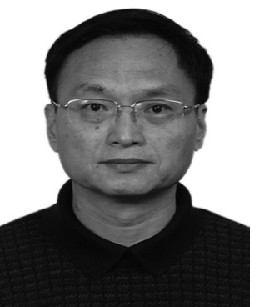}}]{Jun Ma}
received his BSc, MSc, and PhD degrees from Shandong University in China, Ibaraki University, and Kyushu University in Japan, respectively. He is a full professor at Shandong University. He was a senior researcher in the Department of Computer Science at Ibaraki Univsity in 1994 and German GMD and Fraunhofer from the year 1999 to 2003. His research interests include information retrieval, Web data mining, recommendation systems and machine learning. He has published more than 150 International Journal and conference papers, including SIGIR, MM, TOIS and TKDE. He is a member of the ACM and IEEE.
\end{IEEEbiography}
\vspace{-5em}
\begin{IEEEbiography}[{\includegraphics[width=1in,height=1.25in,clip,keepaspectratio]{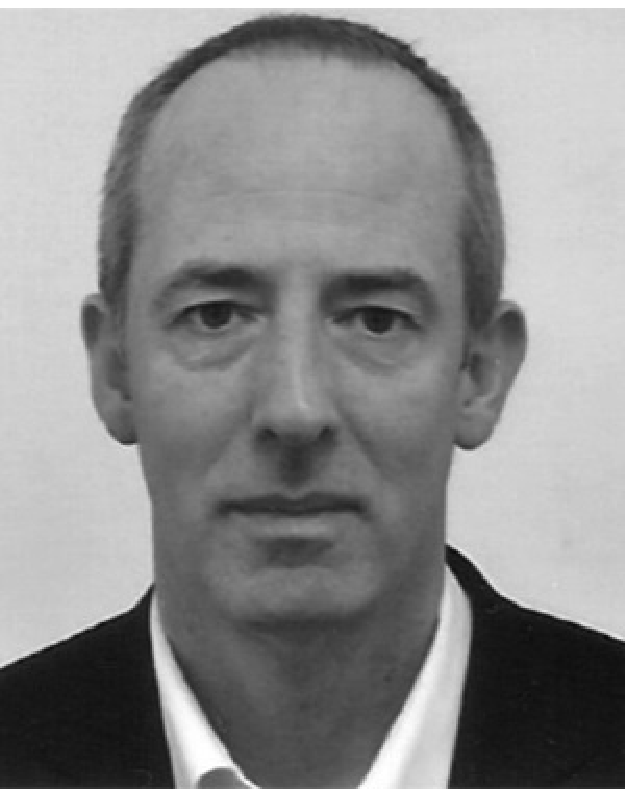}}]{Maarten de Rijke}
received MSc degrees in philosophy and mathematics, and a PhD degree in theoretical computer science. He is a University Professor in Artificial Intelligence and Information Retrieval at the University of Amsterdam. He previously worked as a postdoc at CWI, before becoming a Warwick research fellow at the University of Warwick, United Kingdom. He is the editor-in-chief of ACM Transactions on Information Systems, Springer's Information Retrieval book series, and Foundations and Trends in Information Retrieval.
\end{IEEEbiography}

\end{document}